\documentclass{emulateapj}
\newcommand{\Mpc}{\mbox{ Mpc}}

\newcommand{\eV}{\mbox{ eV}}
\newcommand{\kel}{\mbox{ K}}

\newcommand{\secinv}{\mbox{ s$^{-1}$}}

\newcommand{\hunits}{\mbox{ km s$^{-1}$ Mpc$^{-1}$}}

\newcommand{\bxion}{\bar{x}_i}

\newcommand{\hone}{HI }

\newcommand{\mmin}{m_{\rm min}}
\newcommand{\fcoll}{f_{\rm coll}}

\newcommand{\lya}{Ly$\alpha$ }

\newcommand{\deriv}{{\rm d}}
\newcommand{\bq}{\begin{equation}}
\newcommand{\eq}{\end{equation}}
\newcommand{\bqa}{\begin{eqnarray}}
\newcommand{\eqa}{\end{eqnarray}}
\def\VEV#1{\left\langle #1\right\rangle} 

\begin{document}

\title{Inhomogeneous Helium Reionization and the Equation of State of the Intergalactic Medium}

\author{Steven R.  Furlanetto\altaffilmark{1} \& S.~Peng Oh\altaffilmark{2}}

\altaffiltext{1} {Department of Physics and Astronomy, University of California at Los Angeles; Los Angeles, CA 90095; sfurlane@astro.ucla.edu}

\altaffiltext{2}{Department of Physics, University of California, Santa Barbara, CA 93106;  peng@physics.ucsb.edu}

\begin{abstract}
The temperature of the intergalactic medium (IGM) is set  by the competition between photoheating and adiabatic cooling, which are usually assumed to define a tight equation of state in which the temperature increases monotonically with density.  
We use a semi-analytic model, accurate at low to moderate IGM densities ($\la 5$ times the mean), to show that inhomogeneous reionization can substantially modify these expectations.  Because reionization is driven by biased sources, dense pockets of the IGM are likely to be ionized first.  As a result, voids initially remain cool while dense regions are heated substantially.  However, near the end of reionization, dense regions have already cooled from their initially large temperature while voids have only just been heated.  Thus, near the end of helium reionization the equation of state can invert itself, with the hottest gas inside voids.  The degree to which this happens depends on the magnitude of the density bias during reionization:  if rare, bright sources dominate reionization, so that each ionized region contains a typical volume of the IGM, the equation of state will remain monotonic.  We also show that the distribution of temperatures at a fixed density  has significant scatter and evolves rapidly throughout and even after reionization.  Finally, we show that the observed temperature jump at $z \sim 3.2$ is consistent with the behavior at the end of helium reionization, although it requires a somewhat larger temperature increase than expected.
\end{abstract}
  
\keywords{cosmology: theory -- intergalactic medium -- diffuse radiation}

\section{Introduction} \label{intro}

The two most dramatic events in the history of the intergalactic medium (IGM) are the reionization of hydrogen (by the first generations of galaxies) and helium (by quasars).    These have become key landmarks for both observational and theoretical cosmologists in the past several years.  Evidence for hydrogen reionization comes from a number of directions, none of them clear but all consistent with (possibly extended) reionization at $z \sim 6$--$10$ (see \citealt{fan06-review, furl06-review} for recent reviews).  The clearest evidence for helium reionization comes from \ion{He}{2} \lya forest spectra in the extreme ultraviolet, which seem to show a rapid evolution in the characteristics of the transmission at $z \sim 3$ \citep{davidsen96,anderson99, heap00, smette02, zheng04, shull04, reimers04, zheng04-sdss, reimers05}, although there is also some indirect evidence (see below).

These two reionization epochs are largely responsible for determining the thermal history of the IGM.  Before hydrogen reionization, the neutral IGM cooled adiabatically until the first structures formed, probably reaching temperatures $T \sim 30 \kel$.  X-rays from the first galaxies most likely slowly heated the neutral IGM, to $T \la 1000 \kel$ \citep{oh01, venkatesan01, kuhlen06-21cm, furl06-glob}.  However, hydrogen reionization was a much more dramatic change:  the $\sim 1 \eV$ leftover from each ionizing photon heated the IGM to $\sim 1$--$2 \times 10^4 \kel$ \citep{miralda94, abel99, tittley06}.  The harder photons responsible for helium reionization would have reheated the IGM to similar, or even larger, temperatures \citep{hui97}.  

Once reionization is complete, this heating channel slows dramatically -- because only the relatively small fraction of ions that recombine will couple to the photoionizing background.  The subsequent temperature evolution is determined by a combination of adiabatic heating and/or cooling, photo-heating, and Compton cooling \citep{miralda94, hui97}.\footnote{More precisely, several other factors contribute (see the Appendix of \citealt{hui97} for a summary) but these three are by far the most important.}  The competition between these processes forces the gas temperature to approach an asymptotic form set by the background ionizing spectrum \citep{hui97, hui03}.  Because the magnitude (and indeed sign) of the adiabatic term depends on whether the gas is over- or underdense, the IGM assumes an equation of state $T \approx T_0 (1 + \delta)^{\gamma-1}$, where $\delta$ is the fractional overdensity of the gas element and $\gamma$ is nearly independent of $\delta$ in most models.  

This equation of state inevitably affects many observables of the \lya forest, including the power spectrum of the transmitted flux \citep{croft98, zald01, viel04, mcdonald06}, the optical depth distribution \citep{lidz06-lya}, and the line widths of individual features \citep{schaye99, schaye00, ricotti00, mcdonald01}.  In particular, the observed temperature evolution has been used to constrain the epochs of helium and hydrogen reionization:  most dramatically, the velocity widths of \lya forest absorbers seem to increase sharply at $z \sim 3.2$ \citep{schaye00}, while the equation of state simultaneously flattens \citep{schaye00, ricotti00}.  This may be a result of helium reionization.  If so, the temperature at higher redshifts could provide a clean ``fossil" record of hydrogen reionization:  it appears to be relatively large and so requires that epoch to have occured at $z \la 10$ \citep{theuns02-reion, hui03}.   

The cartoon history described above has been developed from simple uniform reionization models.  These existing treatments have ignored one extremely important facet of reionization:  it is highly inhomogeneous, especially because the sources driving it are themselves highly clustered \citep{barkana04, furl04-bub, furl07-helium}.  As a result, overdense regions in the IGM are ionized before the empty voids (at least on average), and their thermal histories will evolve differently.  This affects the mean equation of state and also adds an element of ``stochasticity" to it, because patches at identical densities can have different reionization histories and hence different temperatures.  To this point, analytic models have ignored this aspect for simplicity, and numerical simulations have not been able to include both the inhomogeneity of reionization (which occurs on large $\ga 10 \Mpc$ scales) while also resolving features of the \lya forest.  The exception is the Monte Carlo, semi-analytic helium reionization model of \citet{gleser05}, who showed that this inhomogeneity significantly affects the equation of state.  However, they used a highly approximate prescription for inhomogeneous reionization.  

In this paper, we present a model that incorporates the latest models of inhomogeneous reionization for calculating the IGM equation of state.  
We apply this model to the helium reionization epoch and show that it has an enormous impact on the thermal properties of the IGM during the best-observed epochs, $z \sim 2$--$4$.  We describe our model for inhomogeneous reionization in \S \ref{inhom} and our model for the subsequent temperature evolution in \S \ref{temphist}.  We examine the resulting temperature distributions in \S \ref{tempdist}, the equation of state in \S \ref{eos}, and the evolution of the IGM temperature in \S \ref{temp-evol}.  We compare our models to observations in \S \ref{comp-obs} and conclude in \S \ref{disc}.

In our numerical calculations, we assume a cosmology with $\Omega_m=0.26$, $\Omega_\Lambda=0.74$, $\Omega_b=0.044$, $H=100 h \hunits$ (with $h=0.74$), $n=0.95$, and $\sigma_8=0.8$, consistent with the most recent measurements \citep{spergel06}.  Unless otherwise specified, we use comoving units for all distances.

\section{Inhomogeneous Reionization} \label{inhom}

Our first goal is to compute the distribution of reionization redshifts for gas elements with mass $m_p$ at a linearized fractional overdensity, extrapolated to the present day, of $\delta_L^0$.  For notational simplicity, we will use $S \equiv \sigma^2(m)$, the variance of the density field smoothed on mass scale $m$, as a proxy for mass in this section; also, $S_p \equiv S(m_p)$.  We will include both helium and hydrogen reionization; for clarity, we will refer to the moment at which ionized hydrogen fills the universe as $z_{\rm H}$ and the moment at which doubly-ionized helium fills the universe as $z_{\rm He}$.\footnote{We assume that helium is singly-ionized at $z_{\rm H}$ as well.}  

For most of our calculations, we will use the excursion set reionization model of \citet{furl04-bub}, which provides an excellent match to simulations of hydrogen reionization, at least when IGM recombinations are relatively uniform \citep{zahn07-comp, mesinger07}.  According to this model, under the reasonable assumption that ionizing photons are produced inside galaxies, the ionization field can be built from the linearized density field by demanding that a region is ionized if and only if it has 
\bq
\zeta \fcoll(\delta_L^0,S) > 1 + \bar{N}_{\rm rec},
\eq
where $\zeta$ is the number of ionizing photons escaping into the IGM per atom (of the appropriate element) inside of galaxies,\footnote{For reference, our ``slow" and ``fast" helium reionization models require $\zeta_{\rm He}=10.2$ and $33.2$ if $z_{\rm He}=3$, and hydrogen reionization requires $\zeta_{\rm H}=40$ if $z_{\rm H}=8$.} 
$\fcoll(\delta_L^0,S)$ is the collapse fraction in a region of density $\delta_L^0$ and mass $S$, and $\bar{N}_{\rm rec}$ is the average cumulative number of recombinations per ionized atom.  Using the excursion set formalism \citep{bond91,lacey93}, this condition can be transformed into the comoving number density of ionized bubbles with masses in the range $(m,\,m_b+dm)$:
\bq
n_b(m,z) = \sqrt{\frac{2}{\pi}} \ \frac{\bar{\rho}}{m^2} \ \left|
  \frac{\deriv \ln \sigma}{\deriv \ln m} \right| \ \frac{B_0(z)}{\sigma(m)} \exp \left[ - \frac{B^2(m,z)}{2 \sigma^2(m)} \right],
\label{eq:dndm}
\eq
where the excursion set barrier has been approximated by the linear function $B(m,z) \approx B_0(z) + B_1(z) S(m)$ and $\bar{\rho}$ is the mean comoving mass density.  Note that, unlike the spherical collapse barrier, this $B(z)$ only extends to a minimum mass scale $m_{b,{\rm min}}$ equal to $\zeta$ times the minimum galaxy mass $\mmin$; smaller ionized bubbles cannot contain any ionizing sources so do not appear in the formalism.

This reionization model has been applied extensively to the epoch of hydrogen reionization, and we will also use it for helium reionization.  There are some key differences that make the  model less 
appropriate for the helium era, but it is nevertheless useful (see \citealt{furl07-helium} for a closer look at these issues).  First, the excursion set formalism divides the universe into completely ionized and completely neutral phases.  This is an excellent approximation for hydrogen reionization, where the mean free paths of ionizing photons are much smaller than the ionized bubbles (assuming that ultraviolet photons drive the process).  It is less good for helium reionization:  because quasars have hard spectra, many of the ionizing photons are able to travel large distances (many Mpc) before interacting wtih a HeII ion.  Thus the boundaries of ionized bubbles will be somewhat ``fuzzy" during helium reionization.  However, the mean free path of most ionizing photons is considerably smaller than the scale of the ionized features throughout most of reionization (e.g., a typical quasar can ionize a region $\sim 10 \Mpc$ across, while $\sim 50\%$ of ionizing photons are absorbed within $\sim 3 \Mpc$).  

Another difference is that helium reionization is mainly driven by relatively rare, bright sources.  As a result, random fluctuations in their abundances play an important role, and because their helium bubbles are so large they will contain a more random mix of dense gas and underdense voids than the bubbles driven by small, clustered sources during hydrogen reionization.  Both these factors imply more stochasticity in the reionization history for a given patch density than in our density-driven model.  
In practice, the early stages of helium reionization are primarily stochastic, but once the ionized fraction exceeds one-half or so, the density-driven component takes over.  We will therefore contrast it with a model in which reionization is completely independent of the local density (but in which the mean ionized fraction evolves the same way).

This distribution allows us to compute the range of ionization histories for our pixels through Bayes' theorem, which in this case says that
\bq
P_{S_p}(z_{\rm ion}>z|\delta_L^0) = \frac{\int_{m_{b,{\rm min}}}^\infty \deriv m \, (m/\bar{\rho}) n_b(m,z) p_{S_p|S}(\delta_L^0| B)}{p_{S_p}(\delta_L^0)}.
\label{eq:bayes}
\eq
Here $P_{S_p}(z_{\rm ion}>z|\delta_L^0)$ is the probability that the element, with $\delta_L^0$ at mass $S_p$, was ionized before $z$, $p_{S_p}(\delta_L^0)$ is the probability that an element of mass $S_p$ has density $\delta_L^0$ (this is simply a gaussian with variance $S_p$), and 
\bqa
p_{S_p|S}(\delta_L^0| B) & = & {1 \over \sqrt{2 \pi (S_p - S)}} \nonumber \\
\times \exp \left[ - \frac{(\delta_L^0 - B)^2}{2(S_p-S)} \right]
\label{eq:pcond}
\eqa
is the conditional probability that an element with mass $S_p$ has density $\delta_L^0$, given that it has density $B$ at mass $S < S_p$.  In other words, the cumulative probability that an element has been ionized before $z$ is the sum of the probabilities that it lies on trajectories passing through the excursion set barrier $B(z)$, normalized by the total abundance of elements of the appropriate final density.  Equation~(\ref{eq:bayes}) can be simplified as
\bq
P_{S_p}(z_{\rm ion}>z|\delta_L^0) = {B_0 \over \sqrt{2 \pi}} \int_0^{S_{b}} {\deriv S \over S^{3/2}} \sqrt{{S_p \over S_p - S}} \exp \left[ - \frac{(\delta S - B S_p)^2}{2 S_p S (S_p-S)} \right].
\label{eq:dpdz}
\eq
Here $S_{b} = \sigma^2(m_{b,{\rm min}})$ is the variance of the density field at the mass scale corresponding to the smallest allowed ionized bubble.  Obviously, if $S_b \ge S_p$ we must have \emph{all} the pixels with $\delta_L^0>B$ ionized.  In reality these scales are not too far apart:  $m_{b,{\rm min}}$ is $\zeta$ times the minimum galaxy mass at the appropriate redshift, while $m_p$ is at the Jeans scale for ionized gas. Their ratio is:
\begin{equation}  
\frac{m_{b,{\rm min}}}{m_{p}}= 1.3 \left( \frac{\zeta}{10} \right) \left( \frac{T_{\rm vir,min}}{2 \times 10^{5} \, {\rm K}} \right)^{3/2} \left( \frac{T_{\rm IGM}}{10^{4} \, {\rm K}} \right)^{3/2}, 
\end{equation}
where $T_{\rm vir, min}$ is the minimum virial temperature of ionizing sources (the fiducial value corresponds to the minimum galaxy mass in photoheated gas).  The similarity between these two scales should come as no surprise, since $m_{b,{\rm min}}$ is definined as the minimum mass which can accrete in a photo-ionized IGM, which is of course intimately related to the Jeans mass. This leads to some ambiguity in our results (see below).

\begin{figure}
\plotone{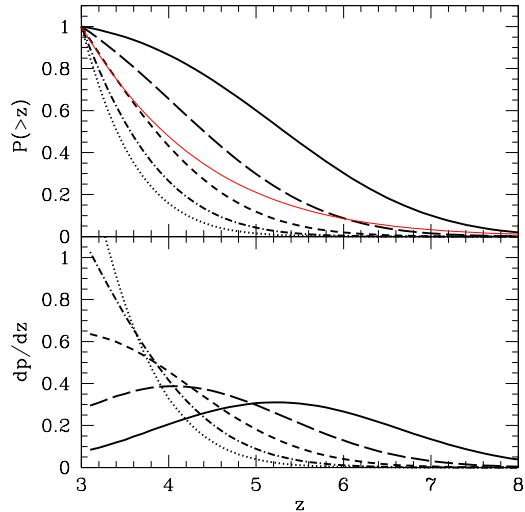}
\caption{Cumulative and differential probability distributions of the helium reionization redshift.  We assume that helium reionization completes at $z_{\rm He}=3$.  In both panels, the thick curves correspond to mass elements with $\delta_L^0=-5,\, -2.5,\, 0,\, 2.5,$ and 5, from left to right.  In the top panel, the thin curve shows the globally-averaged reionization history for our fiducial model.}
\label{fig:dpdz}
\end{figure}

We show the resulting distributions of the local helium reionization redshift for several different $\delta_L^0$ in Figure~\ref{fig:dpdz} (this era has much more significant implications for observations than hydrogen reionization); the upper and lower panels show the cumulative and differential probability distributions, respectively.  We have assumed that reionization ends at $z_{\rm He}=3$.  In each panel, the thick solid, long-dashed, short-dashed, dot-dashed, and dotted curves are for elements with $\delta_L^0=-5,\, -2.5,\, 0,\, 2.5,$ and 5, respectively.  We take $m_p$ to be the Jeans mass of gas in an ionized medium of temperature $10^4 \kel$ at mean density for $z=3$; note that $\sigma(m_p) \approx 2.85$ at this epoch, so these are approximately one- and two-sigma fluctuations at the redshift of helium reionization.

The inhomogeneity of reionization is obvious from Figure~\ref{fig:dpdz}:  even at a fixed density, gas elements can have a wide range of ionization histories.  However, the distributions are strongly dependent on the underlying density, with denser gas much more likely to be ionized at high redshift.  This is simply because such regions are usually found in large-scale overdensities, which are ionized first so long as ionizing sources are more biased than recombining clumps.

\subsection{Dependence on the Reionization Model} \label{pz-reion}

The thin curve in the top panel of Figure~\ref{fig:dpdz} shows the globally-averaged reionization history, $\bxion(z)$.  It mirrors the distribution of reionization redshifts of mean density pockets for much of the evolution, although it has a significantly longer tail (to account for dense gas that is ionized early on).  If $z_{\rm He}$ is fixed, the duration of the reionization era is determined by the rate of change of $\fcoll$.  In our fiducial model, we allow all dark matter halos with $m > \mmin = m_i$ to form quasars, where $m_i$ is the minimum mass for dark matter halos to accrete material in a photoionized medium (with $T \sim 10^4 \kel$); it corresponds to $T_{\rm vir, min} \sim 2 \times 10^5 \kel$.  

\begin{figure}
\plotone{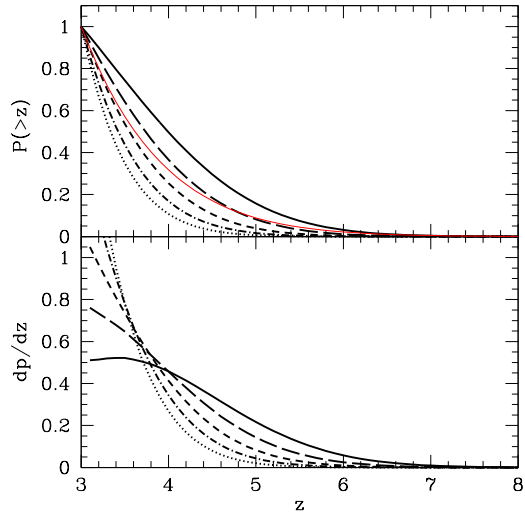}
\caption{As Fig.~\ref{fig:dpdz}, except $\mmin=10 m_i$.}
\label{fig:dpdz_m10}
\end{figure}

This is roughly consistent with other helium reionization models (e.g., \citealt{wyithe03}), but reionization may be less extended if more massive halos dominate the ionizing photon budget.  This can occur in two ways:  by increasing the efficiency of more massive galaxies or by increasing $\mmin$.  As an example, in Figure~\ref{fig:dpdz_m10} we show a model with $\mmin = 10 m_i$.  In this case, $\sim 70\%$ of ionizing photons are produced after $z=4$ (assuming $z_{\rm He}=3$).  This provides a somewhat better match to the reionization history inferred from observations of quasar abundances (in particular, the quasar luminosity function of \citealt{hopkins07}), although uncertainties in the faint end are rather large at high redshifts.  Although the reionization history is faster, note that the qualitative features are the same, and the dependence on density remains substantial.  

As described above, another key component of a helium reionization model is the degree of stochasticity in the process.  The models described above do not include this component, but it is easy to  construct such a model:  for every density, we simply draw $P(>z)$ from the distribution of $\bxion(z)$.  Thus, the histories of \emph{every} gas element in these models tend to mimic those at the mean density of the biased models, but with longer tails toward higher redshift.  We will compare these kinds of models extensively below.

\subsection{Dependence on the IGM Jeans Mass} \label{jeans}

\begin{figure}
\plotone{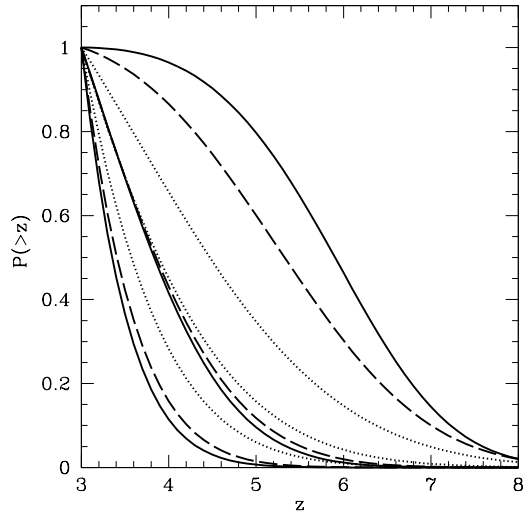}
\caption{Cumulative probability distributions of the helium reionization redshift.  We assume that helium reionization completes at $z_{\rm He}=3$.  The three sets of curves correspond to mass elements with $\delta_L^0=-5,\, 0,$ and 5, from left to right.  Within each set, the solid, dashed, and dotted curves take $T=2 \times 10^4,\,10^4$, and $10^3 \kel$ when evaluating the Jeans mass of the IGM.}
\label{fig:fT}
\end{figure}

Figure~\ref{fig:fT} illustrates the ambiguity in choosing $m_p$.  We show three sets of curves, with $\delta_L^0=-5,\, 0,$ and 5, from left to right.  Within each set, the solid, dashed, and dotted curves take $T=2 \times 10^4,\,10^4$, and $10^3 \kel$ when evaluating the Jeans mass of the IGM.\footnote{These three choices bracket the expected post-reionization IGM temperature (as illustrated in \S \ref{temp-evol} below):  a high temperature would be possible with recent hydrogen reionization and a substantial X-ray background; the $10^3 \kel$ choice is still below the temperature expected with very early hydrogen reionization \citep{hui03}.}  The temperature has the largest effect on high-density regions:  smaller Jeans masses give more ``space" between $S_b$ and $S_p$, allowing trajectories to wander far from the ionization barrier beyond its low-mass edge.  This allows high-density regions to be ionized later than otherwise (by trajectories that begin at low densities and then wander to high densities only at small masses).

The trend is reversed for low-density pixels:  here a smaller Jeans mass translates to somewhat earlier reionization.  Again, this can be understood in terms of the diffusion problem:  if $S_p \gg S_b$, trajectories have an enhanced opportunity to wander from high densities (where they crossed the barrier) to the low-density pixel.  However, the overall difference is smaller in voids than in dense regions, because trajectories must still wander quite far from the ionization barrier to end up as voids at the pixel scale (whereas they can begin just below the ionization barrier and easily wander to high densities).  The two effects roughly cancel at average densities.  Fortunately, this ambiguity does not have a major impact on our final results, which are much more sensitive to the thermal evolution of voids and average-density regions than to overdensities.

\section{The Temperature History} \label{temphist}

Now that we have obtained the distribution of ionization histories, we wish to compute the subsequent temperature evolution.  We will follow a simplified version of \citet{hui97}.

Consider a gas element of fractional overdensity $\delta$ (note that this is the true nonlinear overdensity) and temperature $T$ illuminated by an ionizing background of the form
\bq
J_\nu = J_{{\rm HI},-21} \left( {\nu \over \nu_{\rm HI}} \right)^{-\alpha} \times \left\{
\begin{tabular}{cl}
1 & $\nu_{\rm HI} < \nu < \nu_{\rm HeII}$ \\
$f$ & $\nu_{\rm HeII} < \nu$,
\end{tabular}
\right.
\label{eq:jdefn}
\eq
where $J_{{\rm HI},-21}$ is the angle-averaged specific intensity of the background, in units of $10^{-21}$ erg cm$^{-2}$ s$^{-1}$ Hz$^{-1}$ sr$^{-1}$, $\nu_{\rm HI}$ and $\nu_{\rm HeII}$ are the frequencies corresponding to the \hone and HeII ionization edges, and $f$ is a constant.  We will assume for simplicity that $f=0$ before helium is reionized and $f=1$ afterwards.  We therefore do not evolve the magnitude of the ionizing background with redshift; our results are independent of the amplitude of the background in the highly-ionized limit \citep{hui97}, although our simple treatment will not be completely accurate if the \emph{shape} of the spectrum evolves substantially (which it probably will as quasars begin to dominate over galaxies).  However, this only has a small effect on the calculated thermal history. Our assumed radiation background in equation~(\ref{eq:jdefn}) only affects the species abundance via the assumption of photoionization equilibrium; the heating rates -- which are a complicated function of radiative transfer effects -- are determined separately. 

For simplicity, we will take a fiducial spectrum with $\alpha=1.5$ and $J_{{\rm HI},-21}=1$, similar to that expected from quasars.  This yields photoionization rates, defined via
\bq
\Gamma_i = \int_{\nu_i}^\infty \deriv \nu \, {4 \pi J_\nu \over {\rm h} \nu} \sigma_i,
\label{eq:pion}
\eq
of $\Gamma_{{\rm HI},-12}=2.82 \, (2.83)$, $\Gamma_{{\rm HeI},-12}=4.03 \, (4.29)$, and $\Gamma_{{\rm HeII},-12}=0 \, (0.092)$ before (after) \ion{He}{2} reionization, where $\Gamma = \Gamma_{i,-12} \times 10^{-12} \secinv$.  This spectrum is slightly harder than recent estimates of quasars in the far-ultraviolet \citep{vandenberk01, telfer02} ($\alpha \sim 1.6$--$1.8$, albeit with a wide dispersion), but we use it for consistency with past estimates; spectra in the above range do not appreciably affect our results, although they do slightly decrease the asymptotic temperatures.  Note also that, before quasars dominate the ionizing background, this power law is much too hard for hydrogen ionizing photons, so we do not expect to get the early temperature history precisely correct.  However, we focus on the era around helium reionization, when it is a good approximation.

We define the number density of species $i$ to be $n_i \equiv (1+\delta) \tilde{X}_i \bar{\rho}_b/m_p$, where $\bar{\rho}_b$ is the proper mass density of baryons.  Note that the $\tilde{X}_i$ are the species fractions by mass relative to all baryons (so $\tilde{X}_{\rm HI}$ is \emph{not} the neutral fraction).

The thermal evolution of each element is determined by
\bq
{\deriv T \over \deriv t} = -2 H T + {2 T \over 3(1+\delta)} {\deriv \delta \over \deriv t} - {T \over \sum_i \tilde{X}_i} {\deriv (\sum_i \tilde{X}_i) \over \deriv t} + {2 \over 3 k_B n_t} {\deriv Q \over \deriv t},
\label{eq:Tderiv}
\eq
where $\deriv/\deriv t$ is the Lagrangian derivative.  On the right hand side, the first term accounts for the Hubble expansion, the second for adiabatic cooling or heating from structure formation, the third for the internal energy gained or lost per particle from changing the total particle density, and in the last term $\deriv Q/\deriv t$ is the net heat gain or loss per unit volume from radiation processes (see below) and $n_t$ is the local (total) particle number density.

The second term requires an expression for the growth of the nonlinear density field.  \citet{hui97} used the Zeldovich approximation, along with an analytic estimate for the distribution of the strain tensor, which is probably the best analytic approximation to the density evolution in the quasilinear regime.  However, to maintain computational simplicity, and so as not to introduce another probabilistic element to our code, we will assume that the gas elements evolve following the spherical collapse (or expansion) model.  We map linear densities to nonlinear overdensities via the following fitting formula due to \citet{mo96},
\bq
\delta_L = D(z) \delta_L^0 = \delta_c - {1.35 \over (1+\delta)^{2/3}} - {1.12431 \over \sqrt{1 + \delta}} + {0.78785 \over (1 + \delta)^{0.58661}},
\label{eq:dl-dnl}
\eq
where $D(z)$ is the linear growth function and $\delta_c \approx 1.69$ is the threshold for virialization in the spherical collapse model.  This fitting formula fails at shell-crossing, but we will not be considering such extreme densities.  Under this assumption, we can write
\bq
{\deriv \delta \over \deriv t} = \delta_0^L {\deriv D \over \deriv t} {\deriv \delta \over \deriv \delta_L}.
\label{eq:devol}
\eq

The fourth term in equation~(\ref{eq:Tderiv}) includes a number of radiative heating and cooling processes.  The most important heating mechanism is photoionization itself; each species $i$ contributes a term
\bq
\left. {\deriv Q \over \deriv t} \right|_i = n_i \int_{\nu_i}^\infty \deriv \nu \, (4 \pi J_\nu) \sigma_i \, \left( {{\rm h} \nu - {\rm h} \nu_i \over {\rm h} \nu} \right),
\label{eq:pheat}
\eq
where $\sigma_i$ is the photoionization cross section for species $i$, $\nu_i$ is its ionization threshold, and a Roman ${\rm h}$ denotes Planck's constant (to differentiate it from the Hubble constant).  We use the fits of \citet{verner96} for the photoionization cross sections.  The other relevant mechanisms cool the gas and include Compton cooling off the CMB, recombinations (radiative and, for \ion{He}{2}, dielectronic), collisional ionization, collisional line excitation, and free-free emission.  We use the fits of \citet{hui97} for all of these except Compton cooling (for which we use the exact form) and free-free emission (for which we use the fit in \citealt{theuns98}).  We have verified that our results are unchanged if we use the fits presented in \citet[updated from those in \citealt{cen92}]{theuns98} for all the listed mechanisms.

Because the last two terms in equation~(\ref{eq:Tderiv}) depend on the species abundances, we must supplement them with equations for the evolution of each $\tilde{X}_i$.  These are particularly important during the early stages of reionization, when the $\tilde{X}_i$ change rapidly, along with the photoheating and radiative cooling rates.  However, after this initial phase the abundances rapidly settle into photoionization equilibrium and subsequently evolve quasistatically with the Hubble expansion.  Unfortunately, the thermal evolution during reionization depends on radiative transfer effects and is not well-described by our simple model.  These tend to harden the spectrum \citep{abel99}, because gas near the ionizing sources preferentially absorbs soft photons, leaving more distant gas to be ionized by high-energy photons.  Our procedure instead treats each element in isolation.

For simplicity, rather than solve the $\deriv \tilde{X}_i/\deriv t$ equations explicitly, we initialize our models shortly after reionization by assuming photoionization equilibrium at a fixed temperature $T_i$.  We then assume that the gas remains in photoionization equilibrium at each temperature, density, and redshift.  Beyond the initial phase, this provides excellent agreement with full calculations including the species rate equations but is considerably faster to compute.  

We must then specify $T_i$ for both HI and HeII.  For concreteness, we take $T_i^{\rm H} = 2.4 \times 10^4 \kel$ throughout; this is a rather large value (implicitly assuming significant hardening from radiative transfer, or intrinsically hard reionization sources such as quasars; \citealt{abel99}).  However, our results are quite insensitive to it, because the subsequent cooling is so rapid that memory of the initial post-reionization temperature is erased well before the $z \sim 2$--$4$ regime relevant for us \citep{theuns02-reion, hui03}.  Indeed, we ran our models for smaller values of $T_i^{\rm H}$ and found little difference in the results, except that the pre-helium reionization temperature was slightly smaller (and therefore disagreed even more with measurements; see \S \ref{comp-obs}).  

The choice of $T_i^{\rm He}$ is much more important, because it directly impacts the temperature in the observable redshift range.  Some general arguments do provide bounds on its likely value \citep{miralda94, abel99}.  If the region between the gas element and the ionizing sources is optically thin, the mean excess energy of ionizing photons $E_{\rm thin}$ is the average of the entire spectrum weighted by $\sigma_i \propto E^{-3}$, so $\VEV{E_{\rm thin}}/E_i \approx (\alpha+2)^{-1} \approx 0.3$, where $E_i$ is the ionization potential.  If instead we take the optically thick limit, where all ionizing photons are absorbed, the photon energies are unweighted such that $\VEV{E_{\rm thick}}/E_i \approx (\alpha-1)^{-1} \approx 2$.  In either case, this energy must then be shared with all the IGM baryons  through Coulomb interactions (recall that only $\approx 0.07$ of electrons begin inside of HeII atoms); the net temperature change would then be $\Delta T \approx 0.035 (2/3k_B)  \VEV{E} \sim 0.5$--$3 \times 10^4 \kel$ with our assumed spectrum.  

We will choose $T_i^{\rm He} = 3 \times 10^4 \kel$ for our fiducial model (corresponding to $\Delta T \sim 2 \times 10^4 \kel$ after accounting for the initial temperature), but we will also explore the effects of larger and smaller values.  Note also that we treat $T_i^{\rm He}$ as independent of the underlying density, even though the initial temperature is a (weak) function of the underlying density and $\Delta T$ is actually the constant for a specified spectrum.  We have made this tradeoff in the interests of computational simplicity and transparency.  In fact, we expect both the amplitude and shape of the ionizing background to be density dependent; the radiation field both weakens and hardens as it propagates outwards from the source (e.g., \citealt{bolton04}).  To the extent that voids lie farther from the ionizing sources than dense filaments, they will receive more highly-filtered radiation and hence have larger $\Delta T$. This hardening effect, which we ignore, amplifies the effect we now describe: an inverted equation of state in which voids are substantially hotter than in a homogeneous reionization scenario.  On the other hand, at $z<3$ there is some evidence for a \emph{softer} radiation field in voids than in filaments, which would weaken the effect we describe \citep{shull04}.

We must note that equation~(\ref{eq:Tderiv}) neglects some physical mechanisms that may affect the thermal evolution of certain gas particles.  Most important is shock heating, which dominates the thermal budget of the IGM at low redshifts ($z \la 1$; \citealt{cen99, dave01}) and which cannot be entirely ignored even at higher redshifts \citep{nath01, valageas02, furl04-sh}.  Detailed simulations of large-scale structure shocks show that they typically surround sheets, filaments, and halos but can also extend into low density gas on rare occasions \citep{ryu03, kang05}.  

However, such shocks are likely to have only a small effect on the \lya forest (and diffuse IGM) at $z \ga 2$, because nearly all of the forest -- with the exception of rare, high column density systems -- has densities near or even below the mean.  In this regime, even analytic models that assume photoionization equilibrium and explicitly ignore shocks have had a good deal of success (e.g., \citealt{bi92, bi93, schaye01}).  The reason can be seen in, e.g., Figure~11 of \citet{dave99}, which shows the IGM equation of state measured in a cosmological simulation at a variety of redshifts; this includes large-scale structure shocks but excludes helium reionization, of course.  At $z=3$, the vast majority of gas is confined to the well-defined equation of state.  Shocks, which raise the temperature above that expected from this relation, affect only a small fraction of gas, and virtually all of it has $\Delta \ga 10$.  At $z=2$, more of the gas is in the shocked phase, but again virtually all has $\Delta \ga 10$.  That shocks only affect gas sitting at higher densities than we will study here is not entirely coincidental:  naively, shocks should occur when flows begin converging.  In the cosmic web, this happens when gas turns around to form sheets (at relatively large densities), and in the spherical collapse picture it happens at turnaround ($\Delta=5.55$).  Because our fitting formula in equation~(\ref{eq:dl-dnl}) breaks down at that point anyway, our formalism is not meant to describe the dense gas most subject to shocks.

By $z \sim 1$, shocks become relatively common, and by $z \sim 0$ they dominate the thermal energy of the IGM (again see Fig.~11 in \citealt{dave99}).  In the latter case, they visibly broaden the scatter around the equation of state even at $\Delta<1$, reflecting the growing prevalence of relatively weak shocks in this low density phase.  In this epoch, a proper treatment of shocks would clearly be necessary to understand the IGM.  These weak shocks would probably also exist at $z \ga 2$, but structure formation is much farther behind at that point --  and flows generally have small enough infall speeds that they are unable to shock the gas above the relatively high temperatures already provided by photoionization \citep{furl04-sh}.  In particular, the simulations of \citet{dave99} show that over the redshifts of interest, gravitational shocks cause negligible scatter in the equation of state for $\Delta \la 10$ for the case of hydrogen reionization only. This would only hold with greater effect once the higher sound speeds due to helium reionization are taken into account.  

There are also practical difficulties in incorporating shocks in our approach.  Because the process is stochastic (in that some, but not all, gas elements of a given density are affected), it would require an extra layer of sophistication for our model.  Those few analytic models that do attempt to describe shocks (all of which are untested at high redshifts) do not properly account for this stochasticity at small densities.  For example, the Press-Schechter approach of \citet{furl04-sh} associates shocks with gas that has broken off from the cosmic expansion and thus sets a density threshold below which shocks do not occur.  Thus, we will ignore them in our calculations, although they may add a small amount of scatter (in the form of high temperature gas) to the temperature distributions that we show, particularly at $z=2$; they should be almost negligible at $z>3$.

One additional complication is thermal injection from galaxies, such as galactic-scale winds from star formation.  These hot shells can both clear out gas and heat it.  Fortunately, these winds probably only affect regions near to galaxies, which are relatively dense (and likely to be affected by the nearby ionizing source anyway).  We therefore neglect these in our calculation as well.

\section{The Temperature Distribution} \label{tempdist}

We are now in a position to compute temperature distributions of IGM gas elements.  For a density $\delta_L^0$, we use equation~(\ref{eq:dpdz}) to determine the distribution of \emph{local} reionization redshifts $z_{\rm H}^\star$ and $z_{\rm He}^\star$ (given $z_{\rm H}$ and $z_{\rm He}$ for the Universe at large); we then evolve each packet to a specified redshift to calculate $\deriv p/\deriv T$, the temperature distribution.

There is, however, one more ambiguity in our approach:   each gas element actually has \emph{two} reionization redshifts.  To calculate our temperature distributions, we must therefore specify some relation between $z_{\rm H}^\star$ and $z_{\rm He}^\star$ for a given gas element.  One can imagine two simple scenarios.  First, the two could be completely uncorrelated.  That is, a gas element's redshift of hydrogen reionization has no effect on the helium epoch.  Alternately, the two could be completely correlated:  an element that has $P_{S_p}(z_{\rm ion}>z_{\rm H}^\star)=0.1$ (or that is ionized quite early in the process) will also have $P_{S_p}(z_{\rm ion}>z_{\rm He}^\star)=0.1$ (also early in the process).

There are strong physical grounds to prefer the latter scenario, and we will use it throughout our calculations.  Intuitively, both redshifts should depend on the nearby halo population, and we would expect a region with an overabundance of nearby halos at some high redshift to have an overabundance of quasars later on.  More formally, we are considering a single mass element, so the excursion set trajectory is \emph{fixed} independent of redshift.  The reionization barrier is relatively independent of redshift at a fixed ionized fraction \citep{furl05-charsize}, so we would expect the trajectory to cross the hydrogen and helium barriers at nearly the same ionized fraction.

Of course, this argument is imperfect because the excursion set approach does not properly include stochastic fluctuations in the halo population (which are particularly important for helium reionization). 
In particular, if the number of ionizing sources per discrete bubble is less than a few, Poisson fluctuations in that number are more important than clustering in determining the distribution of bubbles (see discussion in \citealt{furl07-helium}). But it should be a reasonable first guess for our analytic model.  In reality, the choice has relatively little impact on our results, because so long as $z_{\rm He} \ll z_{\rm H}$ the gas elements approach their thermal asymptotes and the precise value of $z_{\rm H}^\star$ is unimportant \citep{hui03}.

\begin{figure}
\plotone{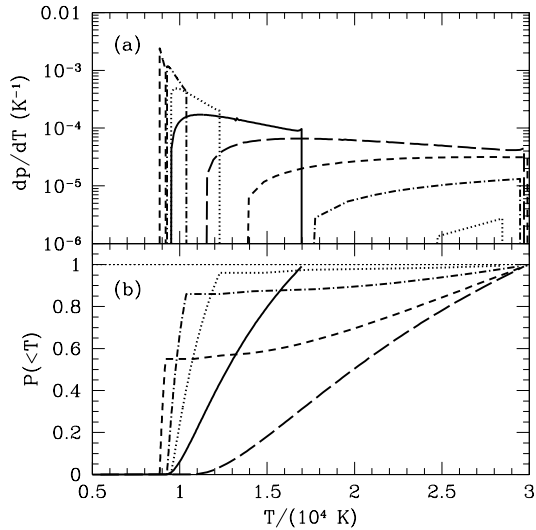}
\caption{Differential and cumulative probability distributions (panels \emph{a} and \emph{b}, respectively) of the IGM temperature for a patch with $\delta_L^0=0$.  The solid, long-dashed, short-dashed, dot-dashed, and dotted curves are for $z=2,\,3,\,4,\,5$, and $6$, respectively.  The plot assumes $z_{\rm H}=8$ and $z_{\rm He}=3$.}
\label{fig:fz}
\end{figure}

Figure~\ref{fig:fz} shows the differential and cumulative probability distributions for the IGM temperature in pixels at the mean density for a variety of redshifts:  $z=2,\,3,\,4,\,5$, and $6$ for the solid, long-dashed, short-dashed, dot-dashed, and dotted curves, respectively.  We have assumed $z_{\rm H}=8$ and $z_{\rm He}=3$.  After helium reionization is complete, at $z \le 3$, the distribution has a relatively simple form, with a relatively flat distribution between $T \sim 10^4 \kel$ and somewhat higher temperatures.  This is a direct result of the broad distribution seen in Figure~\ref{fig:dpdz}:  most elements were ionized relatively early and have cooled to near the thermal asymptote by $z=3$.  Only those elements that were ionized relatively late remain warm.  Both the minimum and maximum of this distribution decrease at $z < z_{\rm He}$ because of adiabatic cooling, until they eventually fill only a small temperature range around the thermal asymptote.

Before helium reionization is complete, however, the distribution is more complex.  This is largely because some elements have undergone helium reionization while others have not.  The sharp spikes at $T \sim 10^4 \kel$ in panel \emph{(a)} represent the gas that has not yet had its helium ionized and so is still approaching the thermal asymptote from hydrogen reionization.  The second component, at much higher temperatures and with a similar shape to the $z=3$ distribution, accounts for elements with $z_{\rm He}^\star>z$.  Note that this part of the distribution does \emph{not} overlap with the thermal asymptote, especially at the higher redshifts.  This gas has undergone helium reionization relatively recently, and cooled since then; hence, the minimum of this component decreases with redshift.  The maximum remains pinned near $T \sim T_i^{\rm He}$ because it represents gas that has just been reionized.

\begin{figure}
\plotone{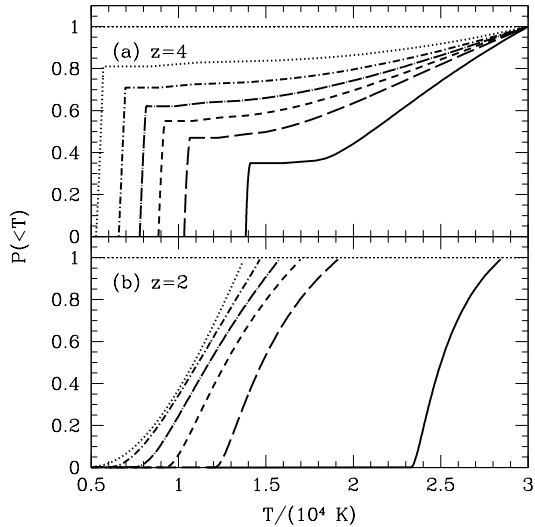}
\caption{Cumulative probability distributions of the IGM temperature for patches at $z=4$ and $2$ (panels \emph{a} and \emph{b}, respectively).  In both panels, the curves take $\delta_L^0=-5,\, -2.5,\, -1,\, 0,\, 1,$ and $2.5$, from left to right.  The plot assumes $z_{\rm H}=8$ and $z_{\rm He}=3$.  In the curves in panel \emph{a}, the sharp rises at $T \la 10^4 \kel$ correspond to regions that have not yet undergone helium reionization.}
\label{fig:fd}
\end{figure}

Of course, these distributions also depend on $\delta_L^0$.  In Figure~\ref{fig:fd}, we show several examples before and after the end of helium reionization (at $z=4$ and $z=2$, respectively).  Again, after $z_{\rm He}$ the distributions are relatively easy to interpret:  regardless of density, the gas has been cooling for a substantial interval and so lies near the thermal asymptote.  Note that  the densest gas (the solid line in Fig.~\ref{fig:fd}\emph{b}) is also the warmest at this stage.  We have already seen that this gas was reionized first, but it is also collapsing, and the accompanying adiabatic heating has raised its temperature well above that of gas at the mean density.  

Again, the distributions are much different before $z_{\rm He}$:  a steep rise at low temperatures representing gas with $z_{\rm He}^\star<4$ and a second, smoothly distributed component at higher temperatures with $z_{\rm He}^\star>4$.  The relative fractions in each component are strong functions of density:  $\la 20\%$ of the most underdense gas has been ionized by $z=4$, but $\ga 60\%$ of dense gas has been.  Again, the characteristic temperature of the cool component increases with density because of adiabatic contraction and expansion.  The distribution of the fully-ionized, high-temperature component is more complex, because the adiabatic heating in dense gas competes with the earlier ionization redshifts.

\section{The Equation of State} \label{eos}

We have seen above that the competition between adiabatic cooling (or heating) and photoheating implies that the IGM temperature depends on the local density, even in the absence of inhomogeneous reionization.  This is usually parameterized by an effective equation of state
\bq
T = T_0 (1+\delta)^{\gamma-1},
\label{eq:eos}
\eq
where $T_0$ is the temperature at the mean density and $\gamma$ is usually taken to be constant over a broad range of density at a given redshift (which turns out to be an excellent approximation for homogeneous reionization; \citealt{hui97}).   In that case, the scatter around this relation is small, so the equation of state has become a primary tool for interpreting the \lya forest.  Usually $T_0$ and $\gamma$ are taken to be free parameters and fit to the observations (although often with priors).

Inhomogeneous reionization modifies the equation of state in two important ways.  First, it produces a substantial scatter in the temperature, even at a fixed density.  This implies that a single equation of state will not be as accurate a representation of the IGM.  Second, the slope $\gamma$ will change because different densities will (on average) be ionized at different times.  We will examine the latter effect first.

\subsection{Inhomogeneous Reionization and $\gamma$} \label{gamma}

\begin{figure*}
\plottwo{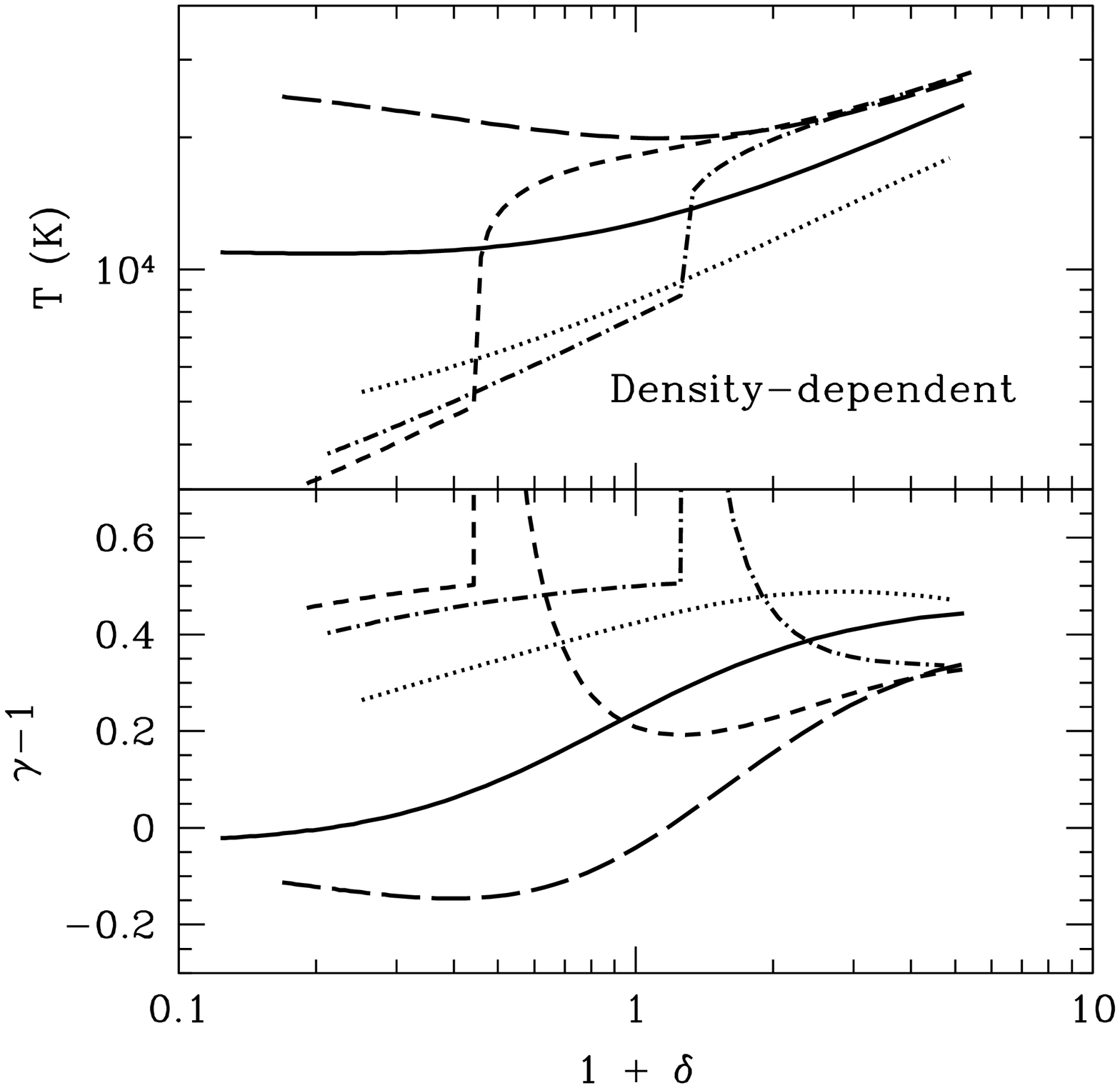}{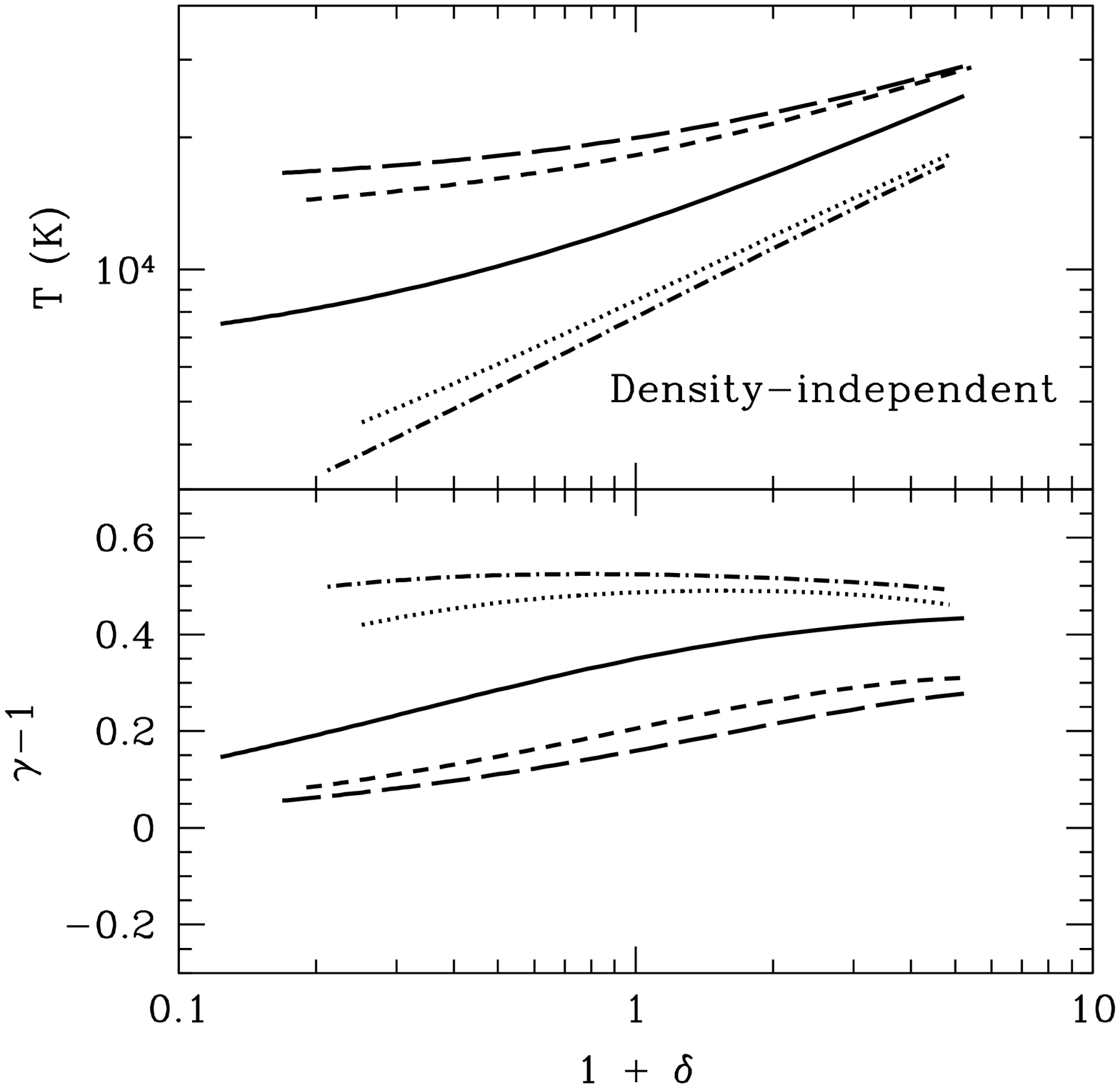}
\caption{Equation of state of the IGM.  The solid, long-dashed, short-dashed, dot-dashed, and dotted curves are for $z=2,\,3,\,3.5,\,4$, and $5$, respectively.   For each density, we plot the corresponding median temperature (\emph{top panels}) and the local logarithmic slope (\emph{bottom panels}).  \emph{Left:}  Excursion set model.  \emph{Right:}  Assumes that the reionization history is independent of local density (see text).  All assume $z_{\rm H}=8$ and $z_{\rm He}=3$, as well as a post-helium reionization temperature of $T_{\rm i}^{\rm He}=30,000$K.}
\label{fig:Tgam}
\end{figure*}

Figure~\ref{fig:Tgam} contrasts the equation of state in the excursion set model (left panels) with a density-independent reionization model (right panels).  We assume $z_{\rm H}=8$, $z_{\rm He}=3$, and $T_i^{\rm He}=3 \times 10^4 \kel$ throughout.  In the top panels, we plot the median temperature as a function of density.\footnote{We choose the median because it better matches the methods used to measure the equation of state directly from \lya forest features (e.g., \citealt{schaye00}).  However, the mean may be more relevant for some applications, like the power spectrum.}  In the bottom panels we show the local logarithmic slope (of the median),
\bq
\gamma - 1 \equiv {\deriv \ln T \over \deriv \ln \delta}.
\label{eq:gammalocal}
\eq
The solid, long-dashed, short-dashed, dot-dashed, and dotted curves are for $z=2,\,3,\,3.5,\,4$, and $5$, respectively.  

First consider density-independent (or stochastic) reionization (at right).  Here the temperature evolution is easy to understand:  $T$ is smallest at the highest redshifts, because helium is still only singly-ionized.  Once helium reionization begins in earnest, the temperature increases everywhere by a factor of several.  Because the post-reionization temperature is independent of local density, this flattens the equation of state, decreasing $\gamma$.  In this particular case, the median temperature distributions at $z=3$ and $z=3.5$ are quite similar, but that is coincidental.  After helium reionization ends, gas begins to cool at all densities, but most significantly in underdense regions.  This translates into an increase in $\gamma$.  At any given redshift, the slope varies only slowly over a large range in density.  Thus, the assumption of an approximately power-law equation of state (as in the homogeneous reionization scenario) is probably still valid. The main difference here from the usual model is the substantial scatter about the median temperature-density relation (see \S \ref{scatter}). 

However, the picture changes completely when the density dependence of reionization is included.  The most obvious change is the sharp increase in the median temperature as a function of density at $z=3.5$--$4$.  This feature is a direct and inevitable result of ``inside-out" reionization, where ionized bubbles grow around galaxies (or quasars), which are highly biased and so include the majority of dense gas elements.  Specifically, the jump occurs at the density at which at least half of the elements have already undergone helium reionization.  Thus it moves to smaller densities as more of the helium is reionized.  

This jump is, of course, accompanied by a discontinuity in the local logarithmic slope.  (This discontinuity is a result of choosing the median temperature rather than an average; the equation of state defined from the average temperature has a much smoother transition but similar endpoints at low and high densities.)  This point also separates two regimes for $\gamma$:  at smaller densities, the curves are fairly steep, because the gas has been cooling since hydrogen reionization.  Thus adiabatic expansion of the voids has steepened the profile.  To the right of the discontinuity, $\gamma - 1 \approx 0.2$; this gas has been ionized relatively recently, with the temperature ``reset" to a density-independent value, so the slope is smaller.

Interestingly, density-dependent reionization has strong effects even after helium reionization is complete (see the $z=2$ and $z=3$ curves):  $\gamma-1 \la 0$ at relatively low densities.\footnote{After $z_{\rm He}$, the mean and median temperatures are very similar, and this negative slope in underdense regions appears in both.} This inverted equation of state implies that voids are hotter than might be expected.  This region even extends to $\delta \sim 0$ at $z=z_{\rm He}$.  This is because underdense regions are ionized later in our model and have had relatively little time to cool.  This trend will of course decay after enough time passes, but it persists at least until $z=2$ in our model, prime territory for \lya forest measurements.  Moreover it is obvious that in this regime it is \emph{not} appropriate to choose a constant value for the logarithmic slope over any substantial density interval.  Instead, the combination of adiabatic cooling, photoheating, and inhomogeneous reionization inevitably leads to a complex equation of state.

\begin{figure*}
\plottwo{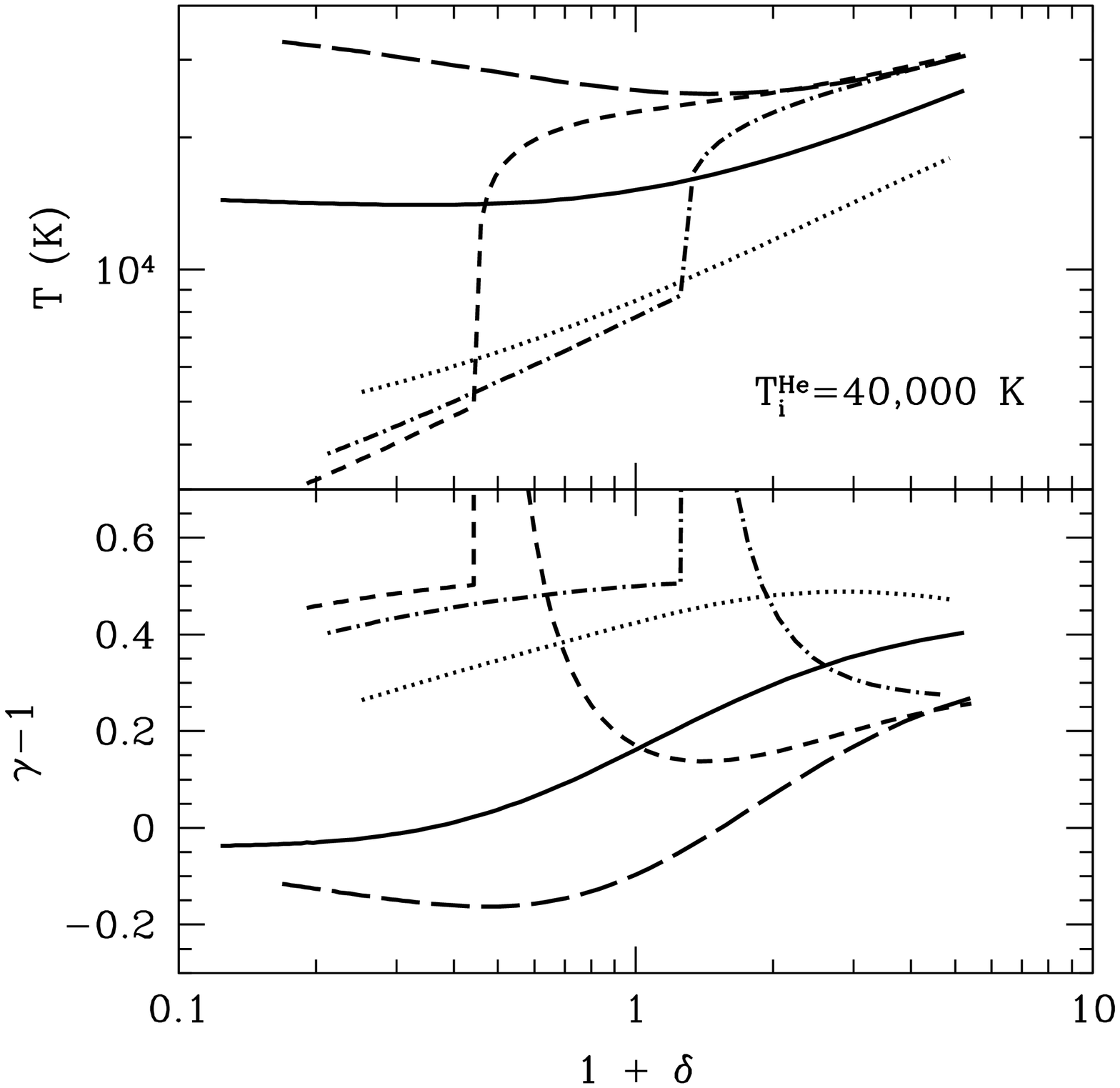}{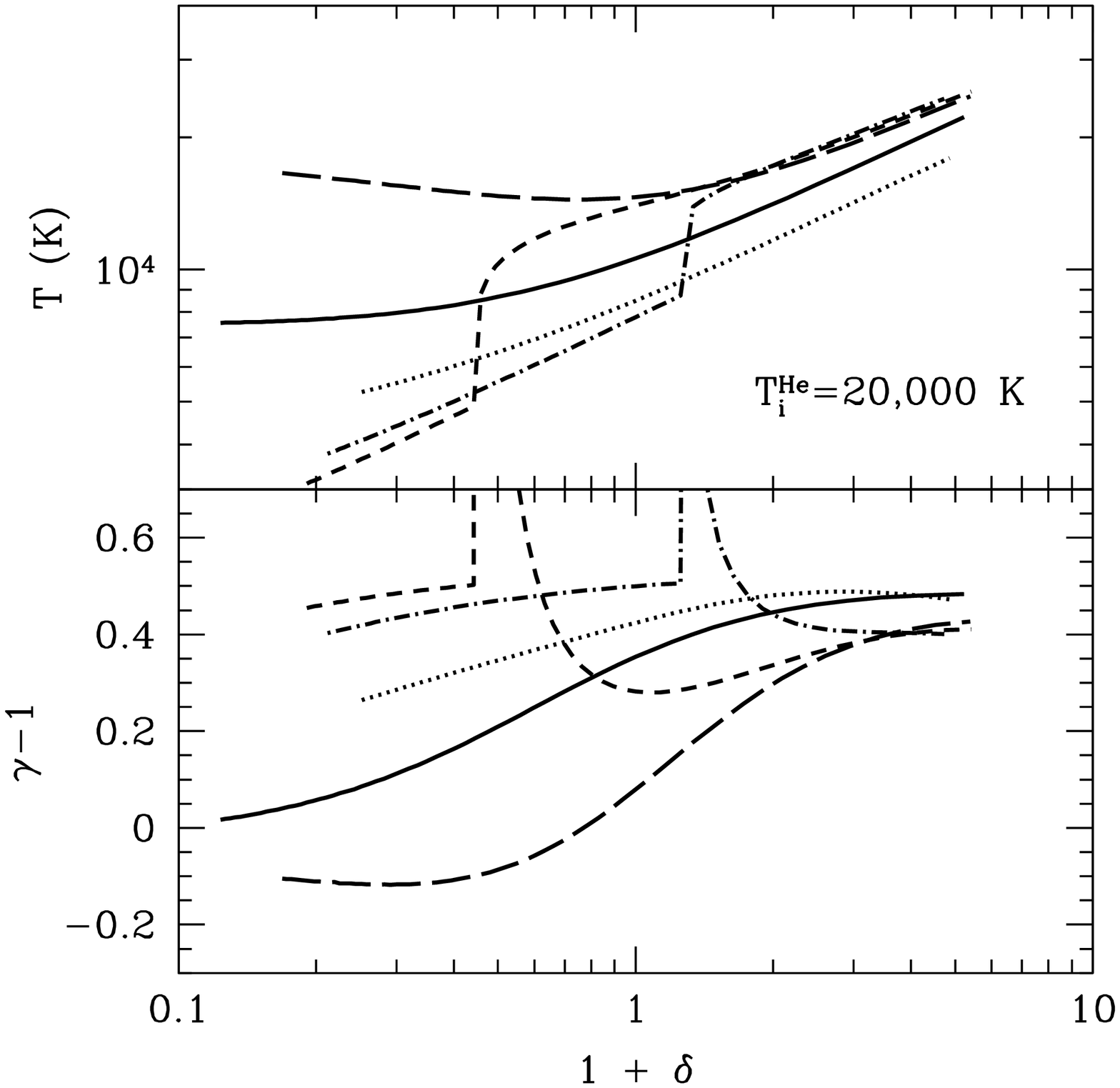}
\caption{Identical to the left panel of Fig.~\ref{fig:Tgam}, except we use $T_i^{\rm He}=4 \ (2) \times 10^4 \kel$ for the left (right) panels.}
\label{fig:Tgam_varTi}
\end{figure*}

Figure~\ref{fig:Tgam_varTi} shows how the equation of state changes with $T_i^{\rm He}$; in the left (right) panels we take $T_i^{\rm He}=4 \, (2) \times 10^4 \kel$.  Not surprisingly, the initial temperature sets the overall scale of the helium reionization break.  On the other hand, the qualitative features are unchanged, and, even in the $T_i^{\rm He}=2 \times 10^4 \kel$ case, we still have sharp breaks in the distribution of the median temperature, an inverted equation of state at low densities, and rapid evolution in the temperature.  Artifacts of helium reionization remain significant even at $z=2$.

\begin{figure*}
\plottwo{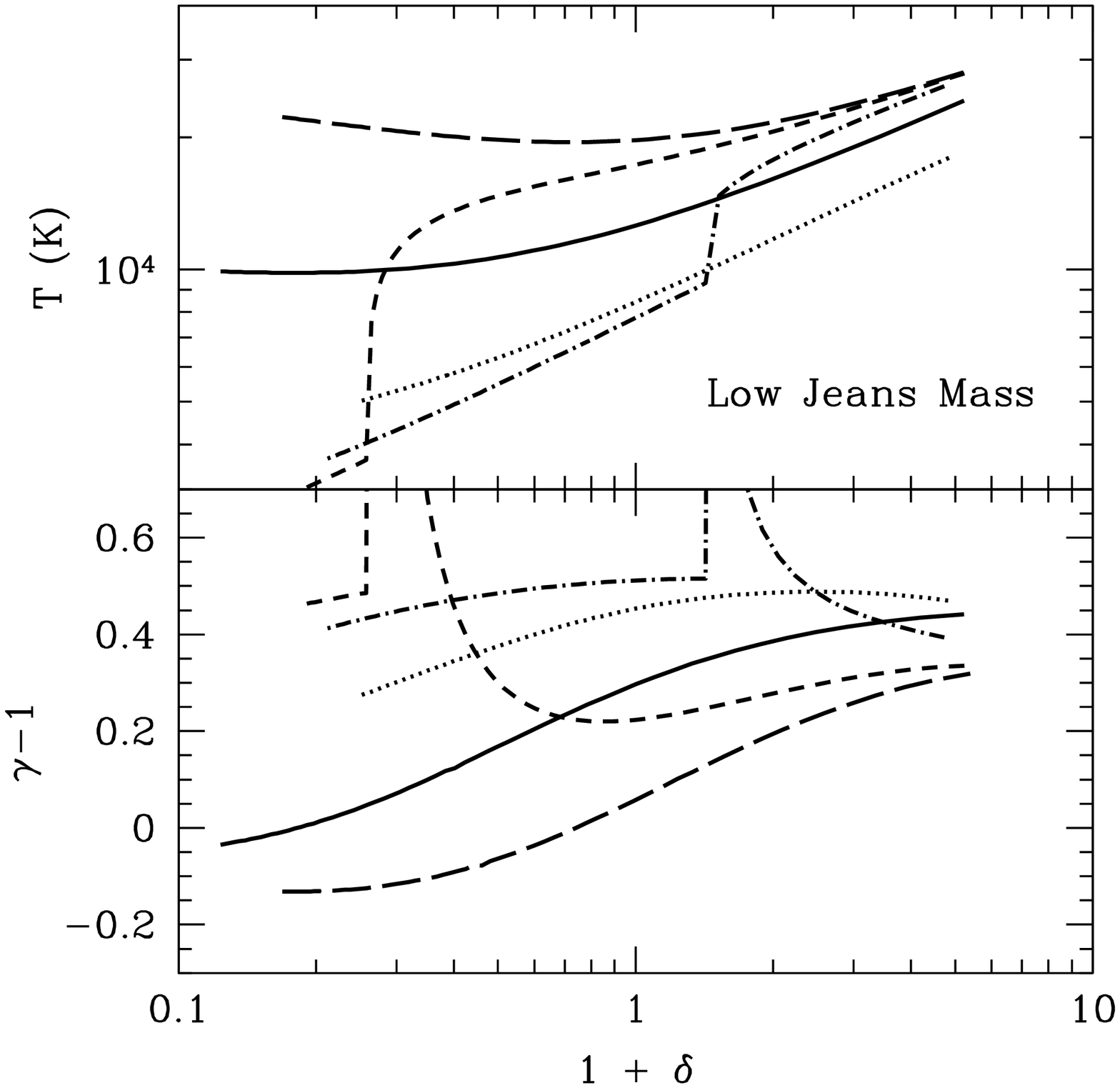}{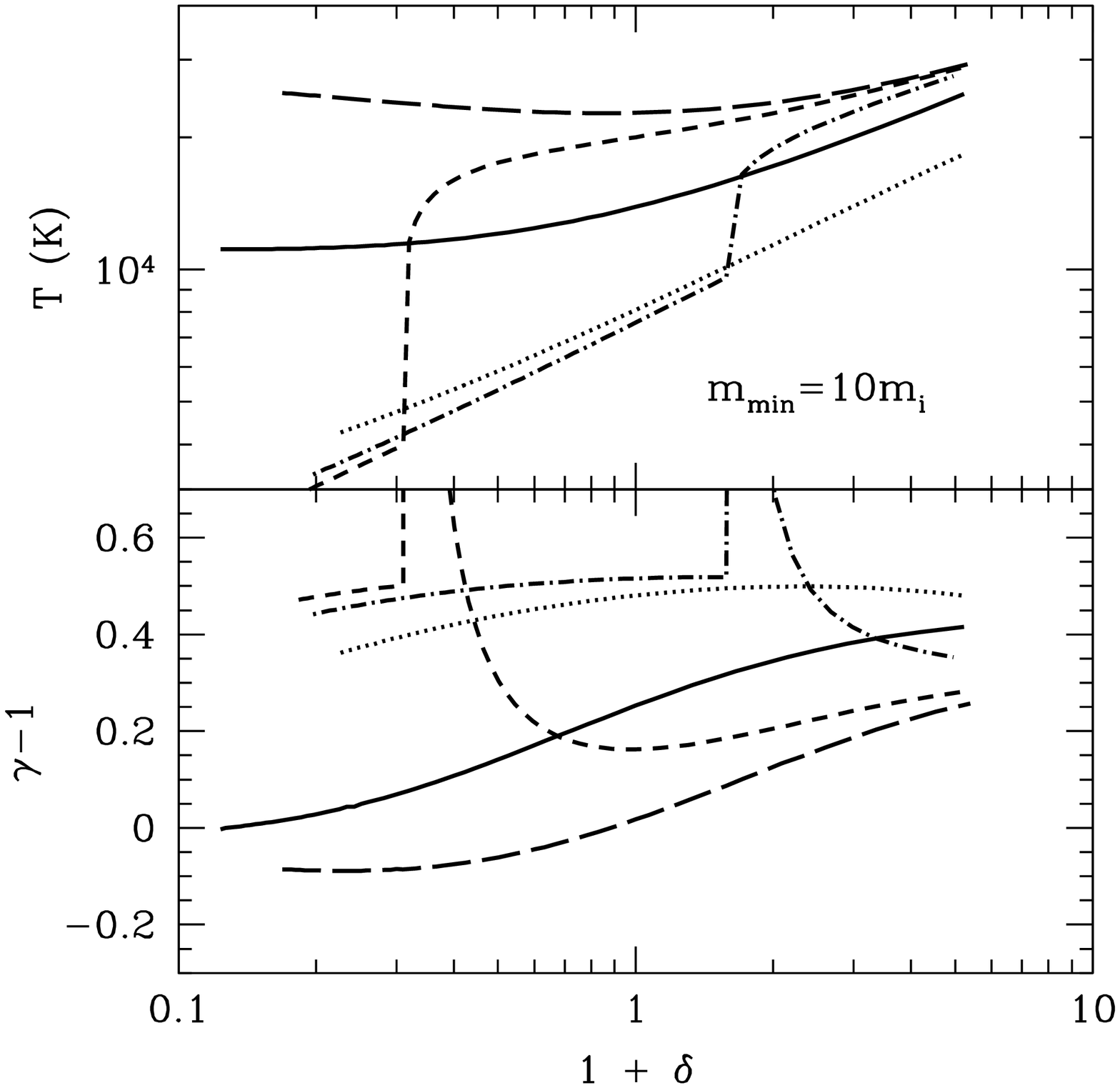}
\caption{\emph{Left:}  Identical to the left panel of Fig.~\ref{fig:Tgam}, except we use $T=10^3 \kel$ to evaluate the Jeans mass of the IGM.  \emph{Right:}  Identical to the left panel of Fig.~\ref{fig:Tgam}, except we use our fast reionization model.  The curves are at $z=2,\,3,\,3.33,\,3.65$, and $4.35$ here; before $z_{\rm He}$, these show the same ionized fractions as in Figure~\ref{fig:Tgam}.}
\label{fig:TdT3}
\end{figure*}

Another uncertainty is that the distribution of $z^\star_{\rm He}$ depends on the assumed mass in each gas element, because that determines how correlated the (small) mass elements are with the (large) ionized bubbles (see Fig.~\ref{fig:fT}).  The left panel of Figure~\ref{fig:TdT3} shows how this affects our equations of state.  Although the detailed relations do change, all of our qualitative points remain valid.  In particular, the sharp jumps in the median temperature as a function of density still appear, although their precise locations change, and $\gamma-1 \la 0$ for small densities, albeit over a somewhat smaller range of $\delta$ and $z$.

The right panel of Figure~\ref{fig:Tgam} shows the evolution in our fast reionization model (with $\mmin=10m_i$, and which probably better matches observations).  In this case, the curves take $z=2,\,3,\,3.33,\,3.65$, and $4.35$; before $z_{\rm He}$, these show the same ionized fractions as in Figure~\ref{fig:Tgam} (0.7, 0.5, and 0.2).  Again, although the locations of the jumps move, the same qualitative features are evident.  The main difference from our fiducial model is simply that the evolution occurs more quickly.

\subsection{Scatter in the Equation of State} \label{scatter}

\begin{figure*}
\plottwo{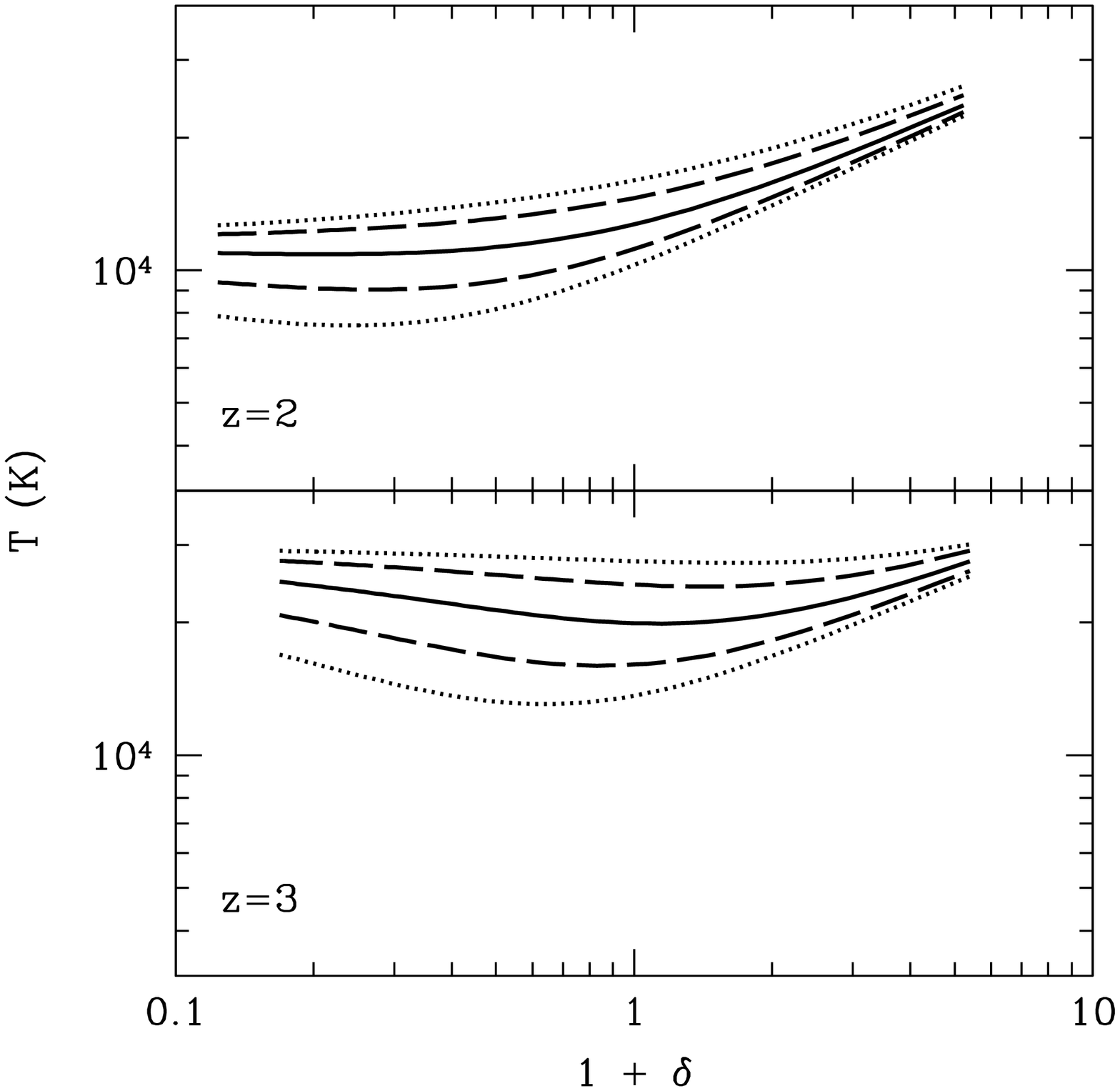}{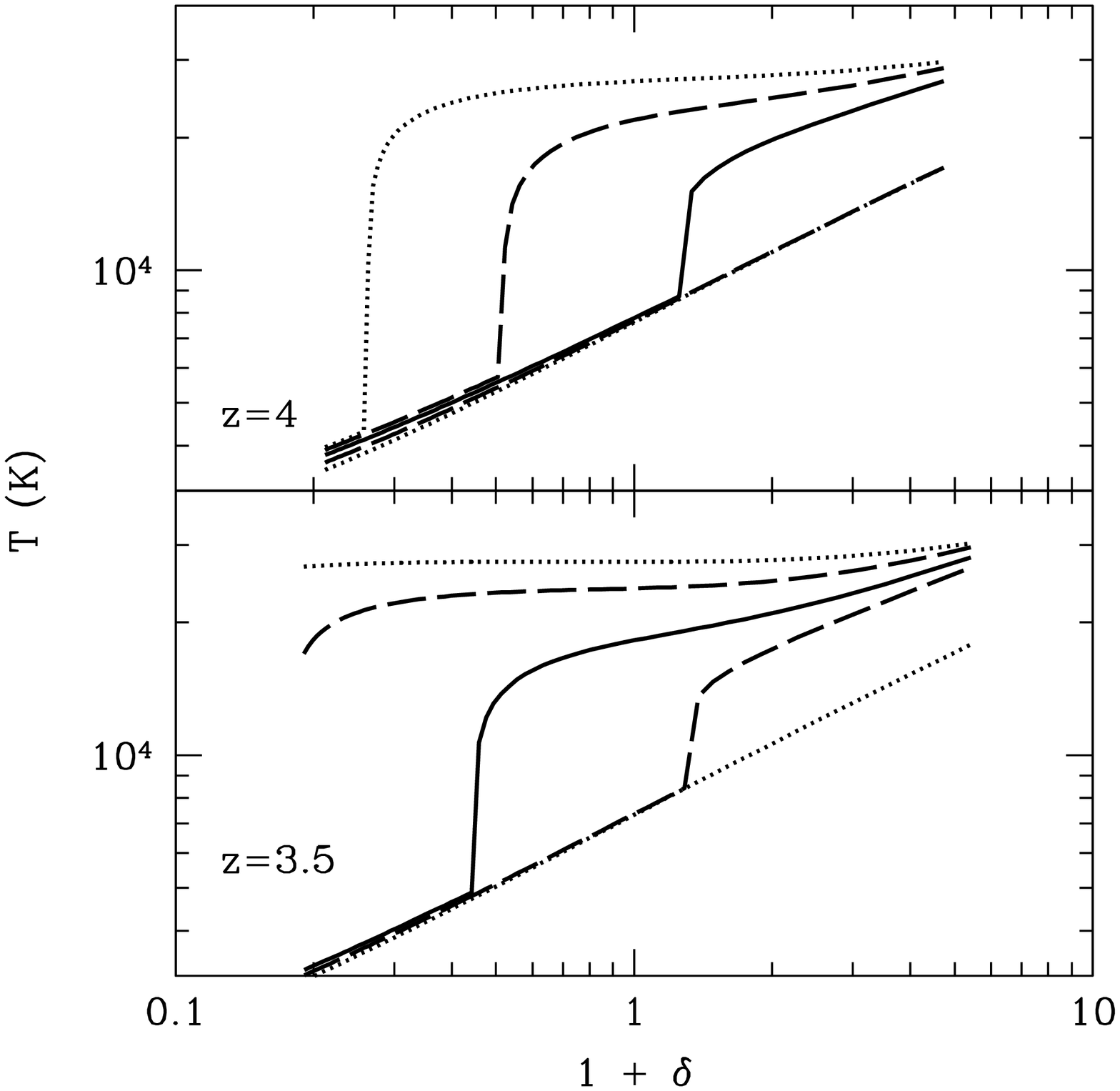}
\caption{Equation of state of the IGM.  In each panel, we show the equation of state taking temperatures from fixed contours of $10\%,\,25\%,\,50\%,\,75\%,$ and $90\%$ in $P(<T|\delta_L^0)$, from bottom to top.  The panels are for $z=2,\,3,\,3.5,$ and $4$, counter-clockwise from the top left.  All assume $z_{\rm H}=8$ and $z_{\rm He}=3$.}
\label{fig:spread}
\end{figure*}

In addition to the change in slope, our model predicts a significant increase in the scatter around the median equation of state (even excluding the modest scatter caused by differences in the tidal fields surrounding the mass elements; see \citealt{hui97}).  We illustrate the scatter at a sequence of redshifts in our fiducial model in Figure~\ref{fig:spread} ($z=2,\,3,\,3.5,$ and $4$, counter-clockwise from top left).  In each panel, the curves show the equations of state for gas with $P(<T|\delta)=0.1,\,0.25,\,0.5,\,0.75,$ and 0.9, from top to bottom.  (Thus the solid curves are identical to those in Fig.~\ref{fig:Tgam}.) As may be seen, the scatter is substantial and may hamper efforts to determine the mean equation of state accurately. 

After helium reionization is complete, the temperature scatter in underdense regions is a factor of two or so, decreasing toward overdense regions.  The scatter is more significant during helium reionization.  Even if most gas elements are on one of the thermal asymptotes, there will be significant outliers in the temperature distribution (differing by up to an order of magnitude from the mean):  in voids, there is a long tail toward high temperatures from the rare gas elements with $z_{\rm He}^\star > z$, and in dense regions there is a long tail toward low temperatures from gas elements with $z_{\rm He}^\star <z$.  Thus, as first pointed out by \citet{hui03}, the \emph{scatter} in the temperature distribution can be used to identify the epoch of helium reionization.  We have shown here that the density range over which this scatter appears will evolve from high to low densities.  This can be used, at least in principle, to trace the history of reionization and to understand how ``inside-out" the process is.

Of course, the duration over which the scatter persists will depend on the duration of reionization:  for example, in the high-mass source model shown in Figure~\ref{fig:dpdz_m10}, the equation of state remains tightly defined until $z \la 3.7$.  The post-reionization scatter in the high-mass source model is smaller, because the range of reionization redshifts (and hence cooling times) is smaller.  However, the differences are not particularly large, and much more careful modeling would be required for such quantitative constraints to be trusted.  In the stochastic model, the scatter is also comparable to that in our fiducial model; because all gas elements more or less track mean density gas, the fractional scatter can be approximately read off from Figure~\ref{fig:spread}.  Of course, in this case there is no density dependence, so there is no jump in the median.

\subsection{Comparison to Previous Work} \label{comp-theor}

To date, the only other theoretical model that allowed for inhomogeneous helium reionization is the Monte Carlo, semi-analytic model of \citet{gleser05}.  They determined the reionization redshift of volume elements probabilistically with the help of a simple biasing prescription (which, like our model, caused dense pixels to be ionized first).  They then followed the subsequent evolution in a similar manner to us, although they allowed the local sources to turn on and off and used the lognormal model for the density evolution (rather than our spherical collapse model).  Their fiducial model had a slightly faster reionization history than our high-mass source model but only took $\Delta T = 10^4 \kel$.

\citet{gleser05} found a similar overall trend for the equation of state (their Fig.~6):  it broke into two branches, with some high-density pixels that are ionized early splitting off to high temperatures and voids staying cool until the end of reionization.  As in our models, the distribution narrows well past reionization.  The most obvious difference is in the actual slope of the equation of state:  \citet{gleser05} only allowed gas to heat by a modest amount during helium reionization, and the equation of state never approached isothermality except over limited ranges of density.  However, Figure~\ref{fig:Tgam_varTi} shows that even in this ``weak" heating case, our model gives a fairly substantial region of near isothermality (in fact even an inverted equation of state).  The source of this difference is likely in the prescription for reionization redshifts, for which \citet{gleser05} used a Monte Carlo implementation with a relatively weak bias.  Their model is thus similar to our density-independent reionization scenario, seen in the right panel of Figure~\ref{fig:Tgam}, where we also find no turnover and only modest flattening.

We conclude that our qualitative conclusions, that helium reionization leads to a complex and evolving equation of state with substantial scatter at low densities, are generic to any ``inside-out" model of reionization.   However, the ``inversion" in the equation of state is more sensitive to the detailed prescription for the ionizing sources.

\section{The Temperature Evolution During Helium Reionization} \label{temp-evol}

We next turn to the evolution of the IGM temperature throughout helium reionization.  Of course, we have already seen that there is a good deal of scatter in $T$, and it depends on the density of the underlying element.  We will follow convention and examine the temperature at a fixed density (the mean); such regions can be identified from their column density in \lya forest spectra \citep{schaye01} and so constitute a well-defined observational sample.  

\begin{figure}
\plotone{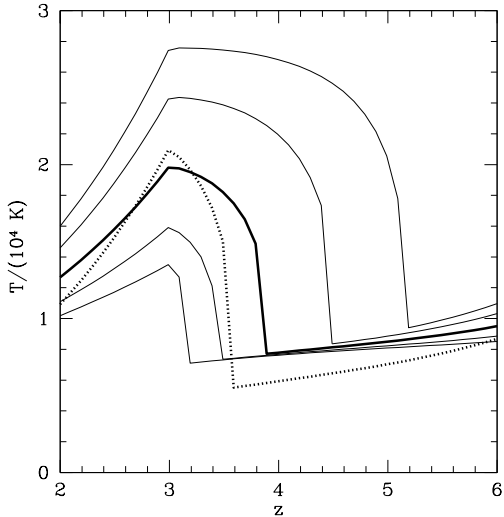}
\caption{Temperature evolution near helium reionization (assumed to be at $z_{\rm He}=3$).  The thick solid curve shows the median temperature for gas elements at the mean density; the thin solid curves show the 10th, 25th, 75th, and 90th percentiles of the temperature distribution.  The dotted curve shows the median temperature for a gas element with $\delta_L^0=-2.5$.}
\label{fig:Tevol_T3}
\end{figure}

Figure~\ref{fig:Tevol_T3} shows the temperature as a function of redshift in the neighborhood of helium reionization (at $z_{\rm He}=3$).  As usual, we take $T_i^{\rm He}=3 \times 10^4 \kel$ here.  Of course, even at a fixed density there is still a range of temperatures in our model; we therefore show the  median temperature at each redshift with the thick solid curve, while the thin solid curves show the 10th, 25th, 75th, and 90th percentiles of the temperature distribution.  

The curves split into three regimes.  At high redshifts, they all lie close together and decrease slowly; this represents gas that still has neutral helium.  Next there is a rapid rise (which, in the case of the thick curve, occurs when $50\%$ of the gas has been ionized).  The initial spike reaches the temperature of the gas that had its helium ionized \emph{first} -- because more recent gas elements are now hotter.  The curves continue to rise, albeit more gently, until $z_{\rm He}$, because they steadily shift toward gas that was ionized later.  Finally, at $z < z_{\rm He}$, the curves decline again, as they simply track a single gas element (ionized halfway through the process for the thick curve) while it cools toward the thermal asymptote.  Note that the scatter decreases again in this regime.

The temperature begins to increase before $z_{\rm He}$, with the initial spike preceding it by a substantial amount:  in the case of the median temperature, it occurs at $z \sim 4$ in this particular model.  The ``reionization redshift" itself is best measured by looking at the minimum temperature (i.e., the last gas element to be ionized).  However, in that case the temperature jump is also the least dramatic, because the statistic jumps from gas with neutral helium to gas that was ionized long before (and has cooled significantly) -- in this case, $\Delta T \sim 6000 \kel$ for the $10$th percentile,\ but $\Delta T \sim 10^4 \kel$ for the median.

The dotted curve in Figure~\ref{fig:Tevol_T3} shows the median temperature for gas elements with $\delta_L^0=-2.5$ (corresponding to a deep void).  Because voids are ionized only at the end of reionization, the temperature increases closer to $z_{\rm He}$, and the jump is somewhat larger (mostly because the initial temperature is smaller).  

\begin{figure}
\plotone{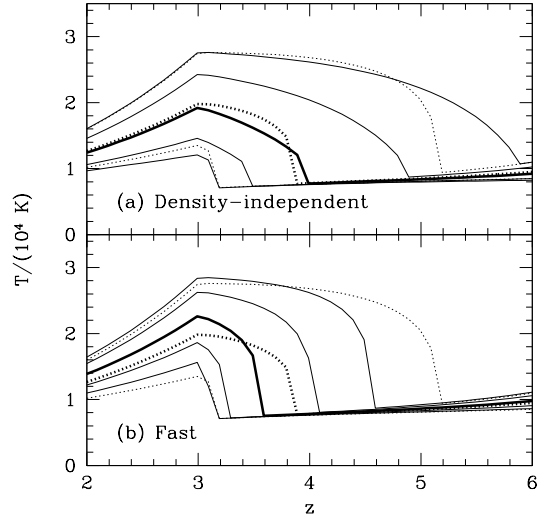}
\caption{Temperature evolution near helium reionization (assumed to be at $z_{\rm He}=3$).  \emph{(a)} The thick solid curve shows the median temperature assuming that the reionization history is independent of density; the thin solid curves show the 10th, 25th, 75th, and 90th percentiles of the temperature distribution.  \emph{(b)} Same, except that the reionization history depends on density but takes $\mmin = 10 m_i$.  In both panels, the dotted curves show the 10th, 50th, and 90th percentile points of the distribution for our fiducial model incorporating inhomogeneous reionization.}
\label{fig:Tevol_params}
\end{figure}

Figure~\ref{fig:Tevol_params} shows how the temperature evolution depends on the details of the reionization model.  In principle, the magnitude of the jump -- which increases for more rapid reionization -- can be used to constrain the duration of reionization, but in practice the differences may be too small to allow robust inference. In panel \emph{(a)}, the solid curves correspond to density-independent reionization; the dotted curves show the fiducial model from Figure~\ref{fig:Tevol_T3} (for the latter, we only show the 10th, 50th, and 90th percentiles).  The differences are quite modest because of course mean-density gas elements have histories typical of the universe as a whole, unless reionization occurs in a pathological manner.  The only obvious difference is in the 90th percentile curve at high redshifts.  With density-dependent reionization, all of the early ionizations occur in dense gas, so these mean density pixels get started later.  Distinguishing reionization driven by density fluctuations from ``stochastic" reionization will require studying a broader range of densities.

The bottom panel is similar, except that the solid curves use our ``fast" reionization model (with $\mmin=10 m_i$) with the usual density-dependent implementation.  Comparison to the dotted curves shows that the the temperature spike occurs at lower redshift (for obvious reasons -- though note that when the 10th percentile is chosen, the delay almost vanishes) and is also somewhat stronger; this is because the median post-reionization gas element has undergone less cooling since it was initially ionized.  The temperature difference is nearly independent of which percentile is chosen.

\begin{figure}
\plotone{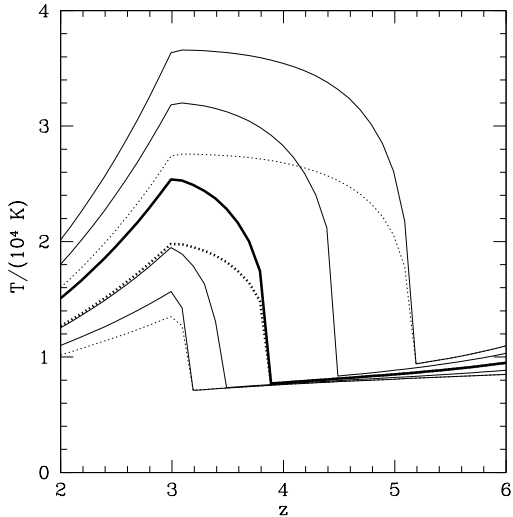}
\caption{Temperature evolution near helium reionization (assumed to be at $z_{\rm He}=3$).  The thick solid curve shows the median temperature for gas elements at the mean density, with $T_i^{\rm He}=4 \times 10^4 \kel$; the thin solid curves show the 10th, 25th, 75th, and 90th percentiles of the temperature distribution.  The dotted curves show the 10th, 50th, and 90th percentile points of the distribution for our fiducial model with $T_i^{\rm He}=3 \times 10^4 \kel$.}
\label{fig:Tevol_T4}
\end{figure}

Figure~\ref{fig:Tevol_T4} shows the temperature evolution if $T_i^{\rm He}=4 \times 10^4 \kel$, again in comparison to the fiducial model.  Of course, this increases the magnitude of the temperature spike without moving its redshift.  However, the change in temperature is now sensitive to the chosen percentile:  for example, the peak of the median changes by $\sim 6000 \kel$ while the peak of the 10th percentile changes by only $\sim 2000 \kel$.  This is because the 10th percentile curve jumps to gas that has already approached the thermal asymptote, washing out $T_i^{\rm He}$.  

\begin{figure}
\plotone{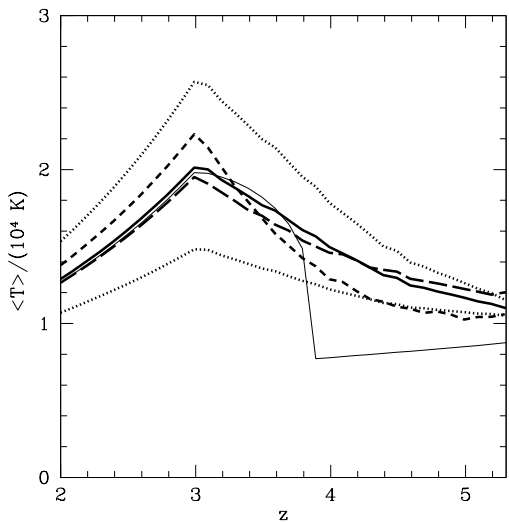}
\caption{Evolution of the mean temperature $\VEV{T}$ near helium reionization (assumed to be at $z_{\rm He}=3$).  The thick solid curve shows our fiducial model.  The dotted curve takes $T_i^{\rm He}=4 \times 10^4 \kel$, the long-dashed curve assumes density-independent reionization, and the short-dashed curves is for our fast reionization model with $\mmin=10m_i$.  For comparison, the thin solid curves shows the median temperature in the fiducial model.}
\label{fig:Tmean}
\end{figure}

Finally, Figure~\ref{fig:Tmean} shows how the \emph{mean} temperature $\VEV{T}$ evolves over this range (again for gas at the mean density).  The thick solid curve shows our fiducial model.  The dotted curves take $T_i^{\rm He}=4 \, (2) \times 10^4 \kel$ (top and bottom, respectively), the long-dashed curve assumes density-independent reionization, and the short-dashed curve is for our fast reionization model with $\mmin=10m_i$.  For comparison, the thin solid curve shows the median temperature in the fiducial model.  Although the mean and the median match almost exactly when $z < z_{\rm He}$, the mean evolves much more smoothly at higher redshifts.  This is simply because the jump in the median occurs when precisely 50\% of the gas has undergone helium reionization, whereas the mean accounts for both cold and reionized gas at all times.  Interestingly, the mean only changes by $\sim 9000 \kel$ in our fiducial model (even though the temperature jump in each gas element is $\ga 2 \times 10^4 \kel$), and it does so over a fairly large redshift interval.

The differences in $\VEV{T}$ between the models are modest; the fast reionization model has slightly higher temperatures after $z_{\rm He}$ (because less cooling has occurred) and slightly smaller temperatures at higher redshifts (because less gas has been ionized).  The density-independent reionization model has a slightly higher temperature at high redshifts, because reionization begins a bit earlier for mean density pixels.  Changing the initial temperature obviously changes $\VEV{T}$, though only by $\sim 60\%$ of the nominal temperature change because of the rapid cooling experienced by elements that are reionized early.  Interestingly, if the energy injection is modest, the observable change in the mean will be quite small.

\section{Comparison to Observations} \label{comp-obs}

The temperature distribution and IGM equation of state are by no means easy to measure, but there have been several attempts in recent years.  These focus on $z \sim 2$--$4$, just when our models predict that the distribution is most interesting.

One approach is to measure the temperature directly through \lya forest features.  Because line widths also depend on the Hubble expansion, turbulence, and peculiar velocities, it is difficult to interpret the velocity width attributed to individual absorbers precisely.  However, one can search for the \emph{minimum} Doppler parameter for systems of a given column density and use that to measure thermal properties \citep{schaye99}.  In detail, \citet{schaye00} compared the observed velocity distribution with a sample drawn from a simulation that did not directly include a model for helium reionization but in which the IGM temperature $T_0$ could be varied.\footnote{See \citet{theuns02-sdss} for a later version that did include uniform helium reionization and matched the measurements reasonably well.}  They then measured $T_0$ from the cutoff in the Doppler width distribution, using it to gauge the combined effects of thermal broadening and Jeans smoothing (the latter was estimated using a suite of numerical simulations).  When compared to our model, this process would ideally measure the \emph{minimum} temperature at each redshift (because their simulation did not include the scatter from inhomogeneous reionization), although with a finite sample it probably actually measures a somewhat higher percentile of the distribution.

Using a sample of high-resolution quasar spectra, \citet{schaye00} found a sharp increase in $T_0$, the fiducial temperature of mean density gas, at $z \sim 3.2$:  from $T_0 \approx 12,500 \kel $ to $T_0 \approx 25,000 \kel$.  Interestingly, the measurements are consistent with a constant temperature at higher redshifts but a rapidly decreasing temperature at lower redshifts (returning to $T_0=12,500 \kel$ at $z=2$); there also appears to be some weak evidence that the jump occurs over a finite ($\Delta z=0.3$) interval.  This is certainly qualitatively consistent with helium reionization ending at $z_{\rm He}=3.2$ \citep{schaye00, theuns02-reion}.  \citet{ricotti00} and \citet{mcdonald01} have made similar measurements.  All are consistent within the errors, although the latter favored a picture without a sharp jump.  However, they did not include Jeans smoothing effects and also chose lines with well-behaved thermal profiles, which may bias the temperature measurement high if there is scatter in the equation of state.

However, there are a number of subtle difficulties.  First, the method of \citet{schaye99} searches for the \emph{minimum} velocity width and attributes it to a combination of thermal broadening and Jeans smoothing.  We would therefore expect that these measurements correspond to a change in the minimum temperature at any given redshift (or at least a low but non-zero percentile).  In that case, we expect a smaller temperature jump than the median over an extremely short redshift interval.  According to our fiducial model, the observed magnitude of the jump ($\Delta T_0 \approx 12,500 \kel$) requires $T_i^{\rm He} \sim 6,\,4,\,$ or $ 3 \times 10^4 \kel$ if the measurement corresponds to the 10th, 25th, or 50th percentile, respectively (or slightly smaller if reionization is faster).  The former two temperatures exceed the limit of what might plausibly be achieved even with extremely optically thick photoheating. Obviously, we must understand the velocity cutoff in the presence of inhomogeneous reionization more precisely, which must be done through a careful calibration against simulations. 

A second difficulty is that the observed pre-reionization temperature is itself rather large, well above naive predictions from models such as ours or from simulations (by $\sim 4000 \kel$), although the errors are large enough that our model cannot be ruled out.  A similar difference occurs in other models (e.g., \citealt{theuns02-reion, hui03}).  This discrepancy can be reduced, but not eliminated, if hydrogen reionization occurred later than $z=8$, or it could be due to turbulence, heating by a substantially harder spectrum from high-redshift quasars, or some other mechanism.  We will not try to model it in more detail here, deferring a closer look at the hydrogen reionization epoch until the future.

A third difficulty is that the observations also cool more quickly than our model at $z < z_{\rm He}$ (albeit again at relatively low confidence), a difference also noted in \citet{gleser05}.  This may be a result of evolution in the ionizing background, which we have ignored \citep{theuns02-reion}.

Similar techniques can also be used to measure the equation of state of the IGM.  So far, the best measurements lie at $z \sim 2$--$4$.  Near the mean density, it appears to be approximately isothermal at $z \sim 3$ \citep{schaye00, ricotti00}, with steepening at both higher and lower redshifts.  Interestingly, the peak in $T_0$ detected by \citet{schaye00} is accompanied by several points with best-fit values $\gamma < 1$, although at less than $1\sigma$ confidence for each one.  This is again consistent with $z_{\rm He}=3.2$ according to our model.  Because inhomogeneous reionization predicts a complex equation of state, with rapid evolution over redshift, it will be interesting to study the equation of state over a broader range in density.

\citet{gleser05} also found that temperature boosts larger than $\Delta T=10^4 \kel$ were necessary to match the \citet{schaye00} measurements, although they worked with the mean temperature, which we have argued is not a good match to the observational methods.  They also do not reproduce the flattening in the equation of state at $z \sim 3$, which requires a larger temperature boost or stronger density dependence during the reionization process.  Overall, it appears that the measurements require relatively hard sources (or substantial IGM filtering) during helium reionization. 

A complementary approach is to use the \lya forest power spectrum of transmitted flux to measure the equation of state.  Such measurements generally show little or no evidence for an evolving equation of state or any substantial change in $T_0$ \citep{theuns00, zald01, viel04, mcdonald05, mcdonald06}, even though they are also most sensitive to gas near the mean density.  However, the errors are large (and in fact the best-fit models often produce unphysical results, unless priors are placed on the thermal history; \citealt{viel04}).  

Figure~\ref{fig:Tmean} helps to resolve this apparent difference with the \citet{schaye00} measurements:  the power spectrum is most likely sensitive to the mean temperature, rather than the details of the distribution.  This evolves much more smoothly than the median and undergoes a substantially smaller increase near helium reionization (at most, it increases by about half the temperature jump of any single gas element).  For example, \citet{zald01} measure $T \sim 2 \times 10^4 \kel$ at $z=3.9$, nearly twice the value estimated by \citet{schaye00}.  However, Figure~\ref{fig:Tmean} shows that, in our fiducial model, we \emph{expect} a difference of $\sim 5000 \kel$ between the mean and median (or minimum), which certainly eases the discrepancy.  The relatively poor redshift resolution available from power spectra will further smooth out this evolution.  More careful modeling is clearly needed, but the power spectrum may in fact be consistent with the \citet{schaye00} measurements.

Power spectrum measurements probably show greater similarity with observations of the evolution in the \ion{H}{1} effective optical depth $\tau_{\rm eff}(z)$, which show a dip in absorption at $z\sim 3.2$ \citep{bernardi03,fauch07}, presumably due to heating from helium reionization (which affects $\tau_{\rm eff}$ through the temperature dependence of the recombination coefficient, $\alpha \propto T^{-0.7}$). The effective optical depth is also more sensitive to the (flux-weighted) mean temperature, and appears to require a jump in IGM temperature at helium reionization by a factor of $\sim 2$ \citep{theuns02-sdss}, even though $\tau_{\rm eff}$ only changes by $\sim 10\%$. However, damping on small scales in the power spectrum, which might be expected after an increase in IGM temperature due to increased Jeans smoothing, is not seen. Reconciling these two measurements will require careful study with numerical simulations. 

\section{Discussion} \label{disc}

We have examined how inhomogeneous reionization affects the equation of state of the IGM.  Because reionization is accompanied by substantial photoheating, it affects the thermal state long after its completion.   We have considered models in which the reionization process proceeds from large-scale overdensities to large-scale voids and in which reionization occurs independently of the underlying density.  In the former case, dense pockets of gas are likely to lie near galaxies (and quasars) and so to be ionized relatively early; voids, on the other hand, are likely to be the last elements to be ionized.  
Our semi-analytic approach ignores shock-heating in the IGM, as well as galactic winds, so it is only accurate at moderate or low overdensities ($\delta \la 4$), where such phenomena are relatively rare, but it does describe the bulk of the gas contributing to the Ly$\alpha$ forest (which sits in photoionization equilibrium).

Hydrogen reionization occurred early enough that the implications for the thermal state in the observable era ($z \la 4$) are relatively modest, and we defer a closer look at fluctuations from that process to the future.  This is primarily because the IGM rapidly approaches a (fairly steep) thermal asymptote, allowing only weak constraints on $z_{\rm H}$ \citep{theuns02-reion, hui03}.  However, \ion{He}{2} reionization at $z \sim 3$--$4$ has a large effect.  

If reionization occurs in a density-independent fashion, the temperature always increases with density, but it is still nearly isothermal during and near the end of reionization.   If underdense gas is ionized last, the equation of state can actually become inverted, with the hottest gas elements in the deepest voids.  In practice, the ``median equation of state," defined with reference to the median temperature at each density,  is reasonably close to isothermal near the end of helium reionization, but with a factor of a few spread in the actual temperatures.  It also develops a break that moves to lower density as the process unfolds (although the break is a result of taking the median; the ``mean" equation of state will be smoother).  Gas elements denser than the break are nearly isothermal (because they have already undergone helium reionization), while those that are less dense have temperatures near the thermal asymptote ($T \la 10^4 \kel$).  

As expected, the temperature increases rapidly near the end of helium reionization, with the magnitude of the jump and its precise timing depending on how it is measured.  The magnitude of the jump depends on $T_i^{\rm He}$, but it is significantly \emph{smaller} than the jump experienced by individual gas elements, because some gas elements must have been ionized at significantly higher redshift and so have cooled toward the new thermal asymptote.  For many types of observations (for example, those sensitive to contours in the temperature distribution), the observed temperature jump will also occur somewhat before reionization is complete.

Measurements of \lya forest lines do show such a temperature jump at $z \approx 3.2$, accompanied by a trend toward a more isothermal equation of state \citep{ricotti00, schaye00}.  This is qualitatively consistent with $z_{\rm He}=3.2$ in our model, but the magnitude of the jump is somewhat higher than one might expect if the \citet{schaye00} method measures the minimum IGM temperature. If so, we would require an extremely hard ionizing background that can produce $T_i^{\rm He} \sim 4 \times 10^4 \kel$.  It is possible, however, that a more realistic assessment of the method with numerical simulations that include inhomogeneous reionization would show that it tends to measure temperatures closer to the median values, in which case a large but not unreasonable initial temperature would be required ($T_i^{\rm He} \sim 3 \times 10^4 \kel$).  

On the other hand, the temperature can also be measured from the \lya forest power spectrum, which shows no evidence for an evolving equation of state and tends to find higher temperatures at $z \sim 4$ \citep{zald01, viel04, mcdonald05, mcdonald06}.  The discrepancy probably arises because the power spectrum method is sensitive to the mean temperature, which evolves more slowly with redshift and has a significantly smaller overall jump than statistics like the median.  Thus inferences about the temperature evolution and equation of state will depend on precisely how a measurement is made (if, for example, the minimum, median, or mean temperature at each density is measured, and which density range is selected).  Careful comparison to observations will probably require detailed simulations of the reionization process.

Of course, our model is far from perfect.  We have ignored fluctuations in the ionizing background (in both its amplitude and spectrum), which will create an even larger spread in the gas temperatures.  Such fluctuations may also depend on the local gas density and introduce biases into the equation of state.   For example, voids lie far from the ionizing sources, so photons that reach them may have been strongly filtered and hence hardened by the IGM; on the other hand, after reionization voids appear to show systematically softer spectra than dense filaments \citep{shull04}.  We have also ignored X-ray heating before a gas element undergoes helium reionization.  In reality, X-rays can travel large distances through the IGM and deposit a fraction of their energy as heat.  This will lift gas off of the thermal asymptote by at least a small amount, especially because such photons will tend to have higher energies than average, and consequently decrease the magnitude of the temperature jump from full reionization.  It may help to reconcile the overall discrepancy of our post-hydrogen reionization temperatures (at $z \ga 4$) with the observations.  For example, if a fraction $\bar{x}_{\rm HeIII}$ of the helium is fully ionized by photons with mean excess energies (above the photoionization edge at $E_{\rm HeII}$) of $\VEV{E}$, we would have
\bq
\Delta T \sim 1500 \kel \left( {\bar{x}_{\rm HeIII} \over 0.1} \right) \left( {\VEV{E} \over E_{\rm HeII}} \right).
\label{eq:simmer}
\eq
Before helium reionization is complete, this ``simmering" background is likely to be quite hard (because the soft photons would be absorbed near the sources), so the heat injection could be substantial.

There are, at least in principle, a variety of other ways to test our model.  One could search for density-dependent temperature fluctuations in the \lya forest lines, for example through correlations between small-scale and large-scale structure in \lya forest lines; the small-scale structure becomes a proxy for temperature through Jeans smoothing.  Such searches have so far been unsuccessful \citep{zald02, theuns02-temp}, but larger samples at a range of redshifts may show a signature.  The substantial increase in the temperature variance during helium reionization will probably be the easiest feature to see (see also \citealt{hui03}), and can provide a diagnostic of when helium reionization occurs and, if the density dependence can be measured, how it proceeds.  Finally, the equation of state also affects the power spectrum of the \lya forest, both directly (because the optical depth $\tau \propto \delta^{2 - 0.7(\gamma-1)}$) and possibly indirectly through spatial fluctuations in the temperature (although \citealt{lai06} have shown that the latter are likely to be small).  

\acknowledgments

We thank C.-A. Faucher-Gigu{\` e}re, L. Hernquist, A. Lidz, A. Loeb, J. Schaye, and M. Zaldarriaga for helpful comments.  SRF acknowledges NSF grant AST-0607470 for support. SPO acknowledges NSF grant AST-0407084 and NASA grant NNG06H95G for support. 

\bibliographystyle{apj}
\bibliography{Ref_21cm,Ref_helium,Ref_2007}

\begin{thebibliography}{68}
\expandafter\ifx\csname natexlab\endcsname\relax\def\natexlab#1{#1}\fi

\bibitem[{{Abel} \& {Haehnelt}(1999)}]{abel99}
{Abel}, T., \& {Haehnelt}, M.~G. 1999, \apjl, 520, L13

\bibitem[{{Anderson} {et~al.}(1999){Anderson}, {Hogan}, {Williams}, \&
  {Carswell}}]{anderson99}
{Anderson}, S.~F., {Hogan}, C.~J., {Williams}, B.~F., \& {Carswell}, R.~F.
  1999, \aj, 117, 56

\bibitem[{{Barkana} \& {Loeb}(2004)}]{barkana04}
{Barkana}, R., \& {Loeb}, A. 2004, \apj, 609, 474

\bibitem[{{Bernardi} {et~al.}(2003){Bernardi}, {Sheth}, {SubbaRao}, {Richards},
  {Burles}, {Connolly}, {Frieman}, {Nichol}, {Schaye}, {Schneider}, {Vanden
  Berk}, {York}, {Brinkmann}, \& {Lamb}}]{bernardi03}
{Bernardi}, M., {Sheth}, R.~K., {SubbaRao}, M., {Richards}, G.~T., {Burles},
  S., {Connolly}, A.~J., {Frieman}, J., {Nichol}, R., {Schaye}, J.,
  {Schneider}, D.~P., {Vanden Berk}, D.~E., {York}, D.~G., {Brinkmann}, J., \&
  {Lamb}, D.~Q. 2003, \aj, 125, 32

\bibitem[{{Bi}(1993)}]{bi93}
{Bi}, H. 1993, \apj, 405, 479

\bibitem[{{Bi} {et~al.}(1992){Bi}, {Boerner}, \& {Chu}}]{bi92}
{Bi}, H.~G., {Boerner}, G., \& {Chu}, Y. 1992, \aap, 266, 1

\bibitem[{{Bolton} {et~al.}(2004){Bolton}, {Meiksin}, \& {White}}]{bolton04}
{Bolton}, J., {Meiksin}, A., \& {White}, M. 2004, \mnras, 348, L43

\bibitem[{{Bond} {et~al.}(1991){Bond}, {Cole}, {Efstathiou}, \&
  {Kaiser}}]{bond91}
{Bond}, J.~R., {Cole}, S., {Efstathiou}, G., \& {Kaiser}, N. 1991, \apj, 379,
  440

\bibitem[{{Cen}(1992)}]{cen92}
{Cen}, R. 1992, \apjs, 78, 341

\bibitem[{{Cen} \& {Ostriker}(1999)}]{cen99}
{Cen}, R., \& {Ostriker}, J.~P. 1999, \apj, 514, 1

\bibitem[{{Croft}(1998)}]{croft98}
{Croft}, R.~A.~C. 1998, in Eighteenth Texas Symposium on Relativistic
  Astrophysics, 664

\bibitem[{{Dav{\'e}} {et~al.}(1999){Dav{\'e}}, {Hernquist}, {Katz}, \&
  {Weinberg}}]{dave99}
{Dav{\'e}}, R., {Hernquist}, L., {Katz}, N., \& {Weinberg}, D.~H. 1999, \apj,
  511, 521

\bibitem[{{Dav{\'e}} {et~al.}(2001)}]{dave01}
{Dav{\'e}}, R., {et~al.} 2001, \apj, 552, 473

\bibitem[{{Davidsen} {et~al.}(1996){Davidsen}, {Kriss}, \& {Wei}}]{davidsen96}
{Davidsen}, A.~F., {Kriss}, G.~A., \& {Wei}, Z. 1996, \nat, 380, 47

\bibitem[{{Fan} {et~al.}(2006){Fan}, {Carilli}, \& {Keating}}]{fan06-review}
{Fan}, X., {Carilli}, C.~L., \& {Keating}, B. 2006, \araa, 44, 415

\bibitem[{{Faucher-Gigu{\` e}re} {et~al.}(2007){Faucher-Gigu{\` e}re},
  {Prochaska}, {Lidz}, {Hernquist}, {Zaldarriaga}, \& {Burles}}]{fauch07}
{Faucher-Gigu{\` e}re}, C.~A., {Prochaska}, J.~X., {Lidz}, A., {Hernquist}, L.,
  {Zaldarriaga}, M., \& {Burles}, S. 2007, submitted to ApJ,
  arXiv.org/0709.2382 [astro-ph]

\bibitem[{{Furlanetto}(2006)}]{furl06-glob}
{Furlanetto}, S.~R. 2006, \mnras, 371, 867

\bibitem[{{Furlanetto} \& {Loeb}(2004)}]{furl04-sh}
{Furlanetto}, S.~R., \& {Loeb}, A. 2004, \apj, 611, 642

\bibitem[{{Furlanetto} {et~al.}(2006{\natexlab{a}}){Furlanetto}, {McQuinn}, \&
  {Hernquist}}]{furl05-charsize}
{Furlanetto}, S.~R., {McQuinn}, M., \& {Hernquist}, L. 2006{\natexlab{a}},
  \mnras, 365, 115

\bibitem[{{Furlanetto} \& {Oh}(2007)}]{furl07-helium}
{Furlanetto}, S.~R., \& {Oh}, S.~P. 2007, \apj, submitted (arXiv.org/0711.1542
  [astro-ph])

\bibitem[{{Furlanetto} {et~al.}(2006{\natexlab{b}}){Furlanetto}, {Oh}, \&
  {Briggs}}]{furl06-review}
{Furlanetto}, S.~R., {Oh}, S.~P., \& {Briggs}, F.~H. 2006{\natexlab{b}},
  \physrep, 433, 181

\bibitem[{{Furlanetto} {et~al.}(2004){Furlanetto}, {Zaldarriaga}, \&
  {Hernquist}}]{furl04-bub}
{Furlanetto}, S.~R., {Zaldarriaga}, M., \& {Hernquist}, L. 2004, \apj, 613, 1

\bibitem[{{Gleser} {et~al.}(2005){Gleser}, {Nusser}, {Benson}, {Ohno}, \&
  {Sugiyama}}]{gleser05}
{Gleser}, L., {Nusser}, A., {Benson}, A.~J., {Ohno}, H., \& {Sugiyama}, N.
  2005, \mnras, 361, 1399

\bibitem[{{Heap} {et~al.}(2000){Heap}, {Williger}, {Smette}, {Hubeny}, {Sahu},
  {Jenkins}, {Tripp}, \& {Winkler}}]{heap00}
{Heap}, S.~R., {Williger}, G.~M., {Smette}, A., {Hubeny}, I., {Sahu}, M.~S.,
  {Jenkins}, E.~B., {Tripp}, T.~M., \& {Winkler}, J.~N. 2000, \apj, 534, 69

\bibitem[{{Hopkins} {et~al.}(2007){Hopkins}, {Richards}, \&
  {Hernquist}}]{hopkins07}
{Hopkins}, P.~F., {Richards}, G.~T., \& {Hernquist}, L. 2007, \apj, 654, 731

\bibitem[{{Hui} \& {Gnedin}(1997)}]{hui97}
{Hui}, L., \& {Gnedin}, N.~Y. 1997, \mnras, 292, 27

\bibitem[{{Hui} \& {Haiman}(2003)}]{hui03}
{Hui}, L., \& {Haiman}, Z. 2003, \apj, 596, 9

\bibitem[{{Kang} {et~al.}(2005){Kang}, {Ryu}, {Cen}, \& {Song}}]{kang05}
{Kang}, H., {Ryu}, D., {Cen}, R., \& {Song}, D. 2005, \apj, 620, 21

\bibitem[{{Kuhlen} {et~al.}(2006){Kuhlen}, {Madau}, \&
  {Montgomery}}]{kuhlen06-21cm}
{Kuhlen}, M., {Madau}, P., \& {Montgomery}, R. 2006, \apjl, 637, L1

\bibitem[{{Lacey} \& {Cole}(1993)}]{lacey93}
{Lacey}, C., \& {Cole}, S. 1993, \mnras, 262, 627

\bibitem[{{Lai} {et~al.}(2006){Lai}, {Lidz}, {Hernquist}, \&
  {Zaldarriaga}}]{lai06}
{Lai}, K., {Lidz}, A., {Hernquist}, L., \& {Zaldarriaga}, M. 2006, \apj, 644,
  61

\bibitem[{{Lidz} {et~al.}(2006){Lidz}, {Heitmann}, {Hui}, {Habib}, {Rauch}, \&
  {Sargent}}]{lidz06-lya}
{Lidz}, A., {Heitmann}, K., {Hui}, L., {Habib}, S., {Rauch}, M., \& {Sargent},
  W.~L.~W. 2006, \apj, 638, 27

\bibitem[{{McDonald} {et~al.}(2001){McDonald}, {Miralda-Escud{\'e}}, {Rauch},
  {Sargent}, {Barlow}, \& {Cen}}]{mcdonald01}
{McDonald}, P., {Miralda-Escud{\'e}}, J., {Rauch}, M., {Sargent}, W.~L.~W.,
  {Barlow}, T.~A., \& {Cen}, R. 2001, \apj, 562, 52

\bibitem[{{McDonald} {et~al.}(2005)}]{mcdonald05}
{McDonald}, P., {et~al.} 2005, \apj, 635, 761

\bibitem[{{McDonald} {et~al.}(2006)}]{mcdonald06}
---. 2006, \apjs, 163, 80

\bibitem[{{Mesinger} \& {Furlanetto}(2007)}]{mesinger07}
{Mesinger}, A., \& {Furlanetto}, S. 2007, \apj, 669, 663

\bibitem[{{Miralda-Escude} \& {Rees}(1994)}]{miralda94}
{Miralda-Escude}, J., \& {Rees}, M.~J. 1994, \mnras, 266, 343

\bibitem[{{Mo} \& {White}(1996)}]{mo96}
{Mo}, H.~J., \& {White}, S.~D.~M. 1996, \mnras, 282, 347

\bibitem[{{Nath} \& {Silk}(2001)}]{nath01}
{Nath}, B.~B., \& {Silk}, J. 2001, \mnras, 327, L5

\bibitem[{{Oh}(2001)}]{oh01}
{Oh}, S.~P. 2001, \apj, 553, 499

\bibitem[{{Reimers} {et~al.}(2005){Reimers}, {Fechner}, {Hagen}, {Jakobsen},
  {Tytler}, \& {Kirkman}}]{reimers05}
{Reimers}, D., {Fechner}, C., {Hagen}, H.-J., {Jakobsen}, P., {Tytler}, D., \&
  {Kirkman}, D. 2005, \aap, 442, 63

\bibitem[{{Reimers} {et~al.}(2004)}]{reimers04}
{Reimers}, D., {et~al.} 2004, to appear in Astrophysics in the Far Ultraviolet:
  Five Years of Discovery with FUSE, ASP Conf. series, eds. G. Sonneborn, W.
  Moos \& B.-G. Andersson (astro-ph/0410588)

\bibitem[{{Ricotti} {et~al.}(2000){Ricotti}, {Gnedin}, \& {Shull}}]{ricotti00}
{Ricotti}, M., {Gnedin}, N.~Y., \& {Shull}, J.~M. 2000, \apj, 534, 41

\bibitem[{{Ryu} {et~al.}(2003){Ryu}, {Kang}, {Hallman}, \& {Jones}}]{ryu03}
{Ryu}, D., {Kang}, H., {Hallman}, E., \& {Jones}, T.~W. 2003, \apj, 593, 599

\bibitem[{{Schaye}(2001)}]{schaye01}
{Schaye}, J. 2001, \apj, 559, 507

\bibitem[{{Schaye} {et~al.}(1999){Schaye}, {Theuns}, {Leonard}, \&
  {Efstathiou}}]{schaye99}
{Schaye}, J., {Theuns}, T., {Leonard}, A., \& {Efstathiou}, G. 1999, \mnras,
  310, 57

\bibitem[{{Schaye} {et~al.}(2000){Schaye}, {Theuns}, {Rauch}, {Efstathiou}, \&
  {Sargent}}]{schaye00}
{Schaye}, J., {Theuns}, T., {Rauch}, M., {Efstathiou}, G., \& {Sargent},
  W.~L.~W. 2000, \mnras, 318, 817

\bibitem[{{Shull} {et~al.}(2004){Shull}, {Tumlinson}, {Giroux}, {Kriss}, \&
  {Reimers}}]{shull04}
{Shull}, J.~M., {Tumlinson}, J., {Giroux}, M.~L., {Kriss}, G.~A., \& {Reimers},
  D. 2004, \apj, 600, 570

\bibitem[{{Smette} {et~al.}(2002){Smette}, {Heap}, {Williger}, {Tripp},
  {Jenkins}, \& {Songaila}}]{smette02}
{Smette}, A., {Heap}, S.~R., {Williger}, G.~M., {Tripp}, T.~M., {Jenkins},
  E.~B., \& {Songaila}, A. 2002, \apj, 564, 542

\bibitem[{{Spergel} {et~al.}(2007){Spergel}, {Bean}, {Dor{\'e}}, {Nolta},
  {Bennett}, {Dunkley}, {Hinshaw}, {Jarosik}, {Komatsu}, {Page}, {Peiris},
  {Verde}, {Halpern}, {Hill}, {Kogut}, {Limon}, {Meyer}, {Odegard}, {Tucker},
  {Weiland}, {Wollack}, \& {Wright}}]{spergel06}
{Spergel}, D.~N., {Bean}, R., {Dor{\'e}}, O., {Nolta}, M.~R., {Bennett}, C.~L.,
  {Dunkley}, J., {Hinshaw}, G., {Jarosik}, N., {Komatsu}, E., {Page}, L.,
  {Peiris}, H.~V., {Verde}, L., {Halpern}, M., {Hill}, R.~S., {Kogut}, A.,
  {Limon}, M., {Meyer}, S.~S., {Odegard}, N., {Tucker}, G.~S., {Weiland},
  J.~L., {Wollack}, E., \& {Wright}, E.~L. 2007, \apjs, 170, 377

\bibitem[{{Telfer} {et~al.}(2002){Telfer}, {Zheng}, {Kriss}, \&
  {Davidsen}}]{telfer02}
{Telfer}, R.~C., {Zheng}, W., {Kriss}, G.~A., \& {Davidsen}, A.~F. 2002, \apj,
  565, 773

\bibitem[{{Theuns} {et~al.}(2002{\natexlab{a}}){Theuns}, {Bernardi}, {Frieman},
  {Hewett}, {Schaye}, {Sheth}, \& {Subbarao}}]{theuns02-sdss}
{Theuns}, T., {Bernardi}, M., {Frieman}, J., {Hewett}, P., {Schaye}, J.,
  {Sheth}, R.~K., \& {Subbarao}, M. 2002{\natexlab{a}}, \apjl, 574, L111

\bibitem[{{Theuns} {et~al.}(1998){Theuns}, {Leonard}, {Efstathiou}, {Pearce},
  \& {Thomas}}]{theuns98}
{Theuns}, T., {Leonard}, A., {Efstathiou}, G., {Pearce}, F.~R., \& {Thomas},
  P.~A. 1998, \mnras, 301, 478

\bibitem[{{Theuns} {et~al.}(2000){Theuns}, {Schaye}, \& {Haehnelt}}]{theuns00}
{Theuns}, T., {Schaye}, J., \& {Haehnelt}, M.~G. 2000, \mnras, 315, 600

\bibitem[{{Theuns} {et~al.}(2002{\natexlab{b}}){Theuns}, {Zaroubi}, {Kim},
  {Tzanavaris}, \& {Carswell}}]{theuns02-temp}
{Theuns}, T., {Zaroubi}, S., {Kim}, T.-S., {Tzanavaris}, P., \& {Carswell},
  R.~F. 2002{\natexlab{b}}, \mnras, 332, 367

\bibitem[{{Theuns} {et~al.}(2002{\natexlab{c}})}]{theuns02-reion}
{Theuns}, T., {et~al.} 2002{\natexlab{c}}, \apjl, 567, L103

\bibitem[{{Tittley} \& {Meiksin}(2006)}]{tittley06}
{Tittley}, E.~R., \& {Meiksin}, A. 2006, submitted to \mnras \
  (astro-ph/0605317)

\bibitem[{{Valageas} {et~al.}(2002){Valageas}, {Schaeffer}, \&
  {Silk}}]{valageas02}
{Valageas}, P., {Schaeffer}, R., \& {Silk}, J. 2002, \aap, 388, 741

\bibitem[{{Vanden Berk} {et~al.}(2001)}]{vandenberk01}
{Vanden Berk}, D.~E., {et~al.} 2001, \aj, 122, 549

\bibitem[{{Venkatesan} {et~al.}(2001){Venkatesan}, {Giroux}, \&
  {Shull}}]{venkatesan01}
{Venkatesan}, A., {Giroux}, M.~L., \& {Shull}, J.~M. 2001, \apj, 563, 1

\bibitem[{{Verner} {et~al.}(1996){Verner}, {Ferland}, {Korista}, \&
  {Yakovlev}}]{verner96}
{Verner}, D.~A., {Ferland}, G.~J., {Korista}, K.~T., \& {Yakovlev}, D.~G. 1996,
  \apj, 465

\bibitem[{{Viel} {et~al.}(2004){Viel}, {Haehnelt}, \& {Springel}}]{viel04}
{Viel}, M., {Haehnelt}, M.~G., \& {Springel}, V. 2004, \mnras, 354, 684

\bibitem[{{Wyithe} \& {Loeb}(2003)}]{wyithe03}
{Wyithe}, J.~S.~B., \& {Loeb}, A. 2003, \apj, 586, 693

\bibitem[{{Zahn} {et~al.}(2007){Zahn}, {Lidz}, {McQuinn}, {Dutta}, {Hernquist},
  {Zaldarriaga}, \& {Furlanetto}}]{zahn07-comp}
{Zahn}, O., {Lidz}, A., {McQuinn}, M., {Dutta}, S., {Hernquist}, L.,
  {Zaldarriaga}, M., \& {Furlanetto}, S.~R. 2007, \apj, 654, 12

\bibitem[{{Zaldarriaga}(2002)}]{zald02}
{Zaldarriaga}, M. 2002, \apj, 564, 153

\bibitem[{{Zaldarriaga} {et~al.}(2001){Zaldarriaga}, {Hui}, \&
  {Tegmark}}]{zald01}
{Zaldarriaga}, M., {Hui}, L., \& {Tegmark}, M. 2001, \apj, 557, 519

\bibitem[{{Zheng} {et~al.}(2004{\natexlab{a}}){Zheng}, {Chiu}, {Anderson},
  {Schneider}, {Hogan}, {York}, {Burles}, \& {Brinkmann}}]{zheng04-sdss}
{Zheng}, W., {Chiu}, K., {Anderson}, S.~F., {Schneider}, D.~P., {Hogan}, C.~J.,
  {York}, D.~G., {Burles}, S., \& {Brinkmann}, J. 2004{\natexlab{a}}, \aj, 127,
  656

\bibitem[{{Zheng} {et~al.}(2004{\natexlab{b}})}]{zheng04}
{Zheng}, W., {et~al.} 2004{\natexlab{b}}, \apj, 605, 631

\end{thebibliography}

\end{document}